\documentclass[iop]{emulateapj}
\usepackage{epsfig,psfig}
\usepackage{auto-pst-pdf}
\usepackage{graphicx, subfigure}
\usepackage{morefloats}
\usepackage{amsmath}
\usepackage{amssymb}
\usepackage{mathrsfs,amssymb}
\usepackage{url}

\newcommand{\swift}{{\it Swift }}

\begin{document}

\title{How long does a burst burst? }

\author{Bin-Bin Zhang\altaffilmark{1}, Bing Zhang\altaffilmark{2}, Kohta Murase\altaffilmark{3}, Valerie Connaughton\altaffilmark{1} and Michael S. Briggs\altaffilmark{1}}

\altaffiltext{1}{Center for Space Plasma and Aeronomic Research (CSPAR), The University
of Alabama in Huntsville, Huntsville, AL 35899, USA; binbin.zhang@uah.edu}
\altaffiltext{2}{Department of Physics and Astronomy, University of Nevada, Las Vegas, Box 454002, 4505 Maryland Parkway, Las Vegas, NV 89154-4002, USA}
\altaffiltext{3}{Hubble Fellow - Institute for Advanced Study, 1 Einstein Dr. Princeton, NJ 08540, USA}

\begin{abstract}
Several gamma-ray bursts (GRBs) last much longer ($\sim$ hours) in $\gamma$-rays than
typical long GRBs ($\sim$ minutes), and recently it was proposed that these ``ultra-long GRBs''
may form a distinct population, probably with a different (e.g. blue supergiant)
progenitor than typical GRBs.
However, \swift observations suggest that many GRBs have extended central engine
activities manifested as flares and internal plateaus in X-rays.
We perform a comprehensive study on a large sample of \swift GRBs with XRT
observations to investigate GRB central engine activity duration and to determine
whether ultra-long GRBs are unusual events. We define burst duration $t_{\rm burst}$
based on both $\gamma$-ray and X-ray light curves rather than using
$\gamma$-ray observations alone. {We find that $t_{\rm burst}$ can be 
reliably measured in 343 GRBs. 
Within this ``good'' sample, 21.9\% GRBs have $t_{\rm burst}\gtrsim10^3$ s and
11.5\% GRBs have $t_{\rm burst}\gtrsim10^4$ s. There is an apparent bimodal 
distribution of $t_{\rm burst}$ in this sample. However, when we consider
an ``undetermined'' sample (304 GRBs) with $t_{\rm burst}$ possibly falling 
in the gap between GRB duration $T_{90}$ and the first X-ray observational 
time, as well as a selection effect against $t_{\rm burst}$ falling into the
first {\swift} orbital ``dead zone'' due to observation constraints, the intrinsic
underlying $t_{\rm burst}$ distribution is consistent with being a single component
distribution. We found that the existing evidence for a separate ultra-long GRB population is 
inconclusive, and further
multi-wavelength observations are needed to draw a firmer conclusion.}
We also discuss the theoretical implications of our results. In particular, the central engine activity duration of GRBs is generally much longer
than the $\gamma$-ray $T_{90}$ duration and it does not even correlate with 
$T_{90}$. It would be premature to make a direct connection between $T_{90}$ 
and the size of the progenitor star. 
 \end{abstract}

\section{Introduction}
A number of GRBs (namely, GRBs 101225A, 111209A, 121027A and the most recent GRB 130925A) were 
found to last much longer ($\sim$ hours instead of tens of seconds) than
typical GRBs \citep{Levan:2013vr,Gendre:2013tr,2013arXiv1310.0313V,2013arXiv1306.1699S}.
Such ``ultra-long" GRBs were also seen historically in BATSE and Konus-Wind data \citep[see, e.g.,][]{1997IAUC.6785....1C,1998tsra.conf..514C, 2002ApJ...570..573G,2002ApJ...567.1028C,2004A&A...427..445N, 2005ApJ...624..880L,Palshin:bl}. {Motivated by such long durations and other multi-wavelength properties (e.g., the faint host galaxy of GRB 101225A and its late time color consistence with SNe II), several groups \citep{Levan:2013vr,Gendre:2013tr} have proposed that
the unusually long durations of these GRBs may point towards a new type of progenitor
stars with much larger radii}, such as blue supergiants \citep{2001ApJ...556L..37M,2013arXiv1307.5061N},
in contrast to the well-accepted compact Wolf-Rayet star progenitor \citep{2006ARA&A..44..507W}. In this scenario, the stellar envelope of a large-radius
massive star would fall back in an extended time scale to fuel the central engine and to
power a relativistic jet. The expected cocoon emission can explain anomalies
in the afterglow data (Nakauchi et al. 2013). If this is the case, then ultra-long
GRBs may form a distinct new population from the traditional short (compact star merger
type) and long (Wolf-Rayet collapsar) GRBs.

However, careful studies based on many more criteria (other than duration alone)
are needed to claim a new population. While the short and long
dichotomy has long been known \citep{1993ApJ...413L.101K}, it was not until
the discoveries of the afterglow, redshift, and host galaxies of
both types of events that a firm claim was made about their distinct
progenitor types. Indeed, based on a dozen multi-wavelength observational criteria
\citep{2009ApJ...703.1696Z}, one was able to establish
robust evidence that long (collapsar/magnetar type) and short (compact star
merger type) GRBs are very different from each other, not only in duration,
but also, more importantly, in their
host galaxy types, specific star formation rate, supernova association,
circumburst medium properties, spectral properties, empirical
correlations, and derived jet opening angles.
Any proposal to claim a new population of GRBs should be performed
in a similar manner. {Even though these multi-wavelength criteria 
are being paid attention to \citep[e.g.][]{Levan:2013vr},
a careful comparative study between the proposed ``ultra-long" GRB 
population and the more classical long GRB population is needed.}

{Interestingly, not all claimed ultra-long GRBs have ultra-long durations
in $\gamma$-rays. Only GRBs 111209A and 130925A 
have an exceedingly long $\gamma$-ray $T_{90}$, i.e. $>10000$ s
\citep[][; Markwardt et al. 2013; Golenetskii et al. 2013]{2011GCN..12663...1G}.
{ GRB 101225A was first measured to have a T$_{90}$ of 1088$\pm$20 s \citep{2010GCN..11500...1P}.
Later studies measured a longer duration of up to 7000 s based on the analysis of gamma-ray data from BAT in subsequent {\it Swift} orbits \citep{2012Natur.482R.120T}.} The gamma-ray duration of GRB 121027A, on the other hand, is only 62.6$\pm$4.8 s
in \swift/BAT band \citep{2012GCN..13910...1B}, which is very typical for long
GRBs. The main { supportive evidence} that GRBs 121027A and 101225A were
included in the ultra-long category was their long-lasting highly-variable X-ray
light curves \citep{Levan:2013vr}. In other words, the ``ultra-long"
durations of GRBs 121027A \citep[``$T_{90}$"$\sim$ 6000s,][]{Levan:2013vr}
and 101225A \citep[``$T_{90}$"$\sim$ 7000s, ][]{Levan:2013vr} { are both
observed in the X-ray band other than being seen in $\gamma$-ray band only}}. In fact, \swift observations over the years have
revealed that the GRB central engine lasts much longer than
indicated by $T_{90}$ \citep{2011CRPhy..12..206Z}, via the manifestation of both X-ray
flares \citep{2005Sci...309.1833B,2006ApJ...642..354Z,2006ApJ...646..351L,2007ApJ...671.1903C,2011MNRAS.410.1064M} and
the so-called ``internal plateaus'' -- X-ray plateaus followed by
an abrupt decay that cannot be interpreted with the external shock model
\citep{2007ApJ...665..599T,2007ApJ...670..565L}. Some authors even suggested that the
entire X-ray afterglow may be of an internal origin powered by central engine
\citep{2007ApJ...658L..75G,Kumar:2008fa,2011ApJ...732...77M}.
The existence of an extended tail emission in most long GRBs was already
hinted from the BATSE data through stacking long GRB light curves \citep{2002ApJ...567.1028C}.
If we believe that GRB duration definition should invoke X-ray data,
then the the duration distribution of GRBs should
be re-analyzed in a systematical manner.

In this paper, we perform a comprehensive study of \swift XRT
data, focusing on the long-term central engine activities in the X-ray light curves,
to address typically how long a burst lasts, and whether the claimed ultra-long GRBs
are special. In \S 2, we propose a new definition, $t_{\rm burst}$, from the physical point of view,
to measure the true time scale of the central engine activity. We also introduce quantitative
observational criteria to measure $t_{\rm burst}$ from data. In \S3, we use the \swift data to
systematically derive $t_{\rm burst}$ and its distribution. We discuss the results
and theoretical implications in \S4.

\section{\lowercase{$t_{\rm burst}$}: Motivation, Definition and Criteria}

Mounting evidence supports the hypothesis that X-ray flares have  the same intrinsic physical origin as $\gamma$-ray pulses, but just have a reduced flux and
peak energy so that they can be below the sensitivity threshold of a $\gamma$-ray
detector (Fig.1 for illustration). For extremely bright X-ray flares, the tips
of the flares can be registered by the $\gamma$-ray detector, and hence, included
in $T_{90}$. Figure 2 gives an example of a GRB (090715B) whose early X-ray flare
as detected by {\em \swift} XRT (red) was also recorded by {\em \swift} BAT (blue),
but the later extended X-ray flares were not. Therefore, $T_{90}$
measurement is not a reliable quantity to describe how long a burst ``bursts''.

In this Paper, we give a physically motivated definition of
the duration of a GRB: {\em The burst duration $t_{\rm burst}$ is an
observable quantity of a GRB, during which the observed ($\gamma$-ray and X-ray) 
emission is dominated by emission from a relativistic jet via an internal
dissipation process (e.g. internal shocks or magnetic dissipation),
not dominated by the afterglow emission from the external shock.}

This definition is different from the traditional $T_{90}$ in that it considers
multi-wavelength signatures in addition to $\gamma$-rays. The rationale of using
such a definition is illustrated in a cartoon picture in Fig.1. The GRB central
engine continuously ejects energy but generally with a reduced power as a function
of time. The peak energy of the spectrum $E_p$ is positively correlated to its
luminosity \citep[e.g.][]{2012ApJ...756..112L}, so it decreases with time. At a certain epoch
($\sim T_{90}$), the signal drops out from the $\gamma$-ray band, but it still
continues in the X-ray band. On the other hand, the afterglow component sets
in early on, peaking at $t_{ag,p}$ and decays with time. It is initially over-shone
by the internal-origin X-ray component (X-ray flares and plateaus). Since the
decay of internal emission is typically very steep, the afterglow component will
eventually show up. The X-ray light curve
therefore displays a steep-to-shallow transition when the external shock component
emerges. In principle, the central engine can activate again to power bright internal
emission to outshine the afterglow component again later. So a secure {\em lower limit}
of the central engine activity time should be defined by the {\em last} observed
steep-to-shallow transition, and this is our definition of $t_{\rm burst}$.

Such a definition is however not easy to quantify. This is because in order
to claim an internal origin of X-ray emission, theoretical modeling is
needed to exclude an external shock origin of the observed flux. The standard
external shock afterglow model (e.g. Gao et al. 2013 for a review) generally predicts
broken-power-law light curves. The steepest decay can be achieved when
the blastwave enters a void, during which emission is powered by the high-latitude
emission\citep{zhangbb07,zhangbb09}. The decay slope in this regime is $\alpha = 2 + \beta$ \citep[convention
$F_\nu \propto t^{-\alpha} \nu^{-\beta}$;][]{2000ApJ...541L..51K}, which is
typically smaller than 3. Due to the equal-arrival-time surface effect, any
variability in external shock emission should satisfy $\Delta t / t \geq 1$,
where $\Delta t$ and $t$ are the variability time scale and the epoch of
observation, respectively \citep[e.g.][]{2005ApJ...631..429I}. As a result, rapid variabilities
with $\Delta t/t \ll 1$ (as observed in X-ray flares) and any steep decay with
slope steeper than -3 (as observed in ``internal X-ray plateaus'') are deemed
as due to an internal origin.

We therefore adopt the following procedure to define $t_{\rm burst}$ of a GRB:
1) Calculate $T_{90}$ for the \swift/BAT light curve; 2) Fit the \swift/XRT light curve as a multi-segment
broken power-law; 3) Identify the steep-to-shallow transitions in the light curve,
and record the decay slope before the transition; 4) Identify the {\em last} transition
with pre-break slope steeper than -3, and record the transition time\footnote{\cite{2013ApJ...763...15Q} 
and \cite{2013arXiv1302.4876P} also discussed
GRB central engine time scale using the X-ray flare data. They used the peak
of the last X-ray flare to define $t_{\rm burst}$.}. The burst 
duration $t_{\rm burst}$ is defined as the maximum of this transition time
and $T_{90}$ of $\gamma$-ray emission\footnote{Here it is assumed that
emission during $T_{90}$ is due to internal emission powered by central engine
activity. This hypothesis is valid for most high-luminosity GRBs, which is
supported by the observed rapid variability of the gamma-ray light curves
as well as the X-ray follow-up steep decay phase following $\gamma$-ray
emission.}.

Notice that this method identifies only the X-ray emission that {\em must} be of an internal
origin, but may not necessarily catch the full duration of internal emission if some
internal-origin emission does not show such a steep decline
\citep[e.g.][]{2007ApJ...658L..75G,Kumar:2008fa,2011ApJ...732...77M}.
Therefore, we may typically regard $t_{\rm burst}$ as the {\em lower limit} of
GRB central engine activity.

\section{Observed \lowercase{$t_{\rm burst}$} distribution}

{As of 2014 January 22nd, 712 GRBs have X-ray afterglows detected by \swift/XRT. }
 All the XRT light curves are directly taken from the \swift/XRT team website\footnote{http://www.\swift.ac.uk/xrt\_curves/} \citep{Evans:2009kx} at the UK \swift Science Data Centre (UKSSDC), which were processed using HEASOFT v6.12. Several example light curves are presented in Fig.3, including the four ultra-long GRBs and some typical GRBs with canonical X-ray light curve behavior. One can see that the central engine activity usually lasts much longer than T90.

{In order to measure $t_{\rm burst}$, we use only well-sampled XRT light curves with late-time observations. We select a ``good'' sample based on the following criteria: (1) The X-ray light curve must have at least 6 data points, excluding upper limits; (2) The X-ray light curve has at least one steep-to-shallow transitions (with the steeper slope $<$ -3); or (3) if the X-ray light curve has no steep-to-shallow transition, the starting time of XRT observation, $T_{X,0}$, is smaller than $T_{90}$. For this latter case, we take $T_{90}$ as $t_{burst}$. Our final good sample consists of 343 GRBs (Table 1). This ``good'' sample, despite having robust measurements of $t_{\rm burst}$, is incomplete. A good fraction of GRBs (consisting of 304 GRBs), which we define as the ``undetermined'' sample, have at least 6 data points in the light curves, do not have a required steep-to-shallow transition (with steeper slope $<$ -3), but have an observational gap between $T_{90}$ and $T_{X,0}$. The $t_{\rm burst}$ of these GRBs likely fall into the gap between $T_{90}$ and $T_{X,0}$, but are not included in the ``good'' sample. Therefore the ``good'' sample is biased against GRBs with a short $t_{\rm burst}$.}

The essential part of measuring $t_{\rm burst}$ is to identify a shallower break feature in the late segments of the X-ray light curve.
This is tricky, since late time X-ray data sometimes have too few photons, or the entire light curves lack time coverage\footnote{A low Earth orbit satellite is subject to Earth occultation, which would affect detections of long-lived emission. This effect is discussed more in \S 4.}. {To maximize the use of the observational data, we apply a
multivariate adaptive regression splines technique \citep[e.g.,][]{Friedman:1991} to the observed light curves in the logarithmic scale, which can automatically detect and optimize breaks\footnote{{Our results are consistent with the fitting results obtained by \cite{Evans:2009kx} (see, e.g, \url{http://www.swift.ac.uk/xrt\_live\_cat/}), but we do not exclude the steep decay and flare phases, which are essential to measure $t_{\rm burst}$}}.} By measuring the decay slope before the break, one can judge whether the pre-break emission is internal, and hence, to measure $t_{\rm burst}$. Figure 4 shows several examples of such measurements. {In several cases (e.g, GRBs 130925A, 121027A, 111209A, 090715B and 051117A), such a break is clearly identified so that $t_{\rm burst}$ is measured. In a few cases (e.g, GRB 140102A), such a break is not identified, but there is overlap between $\gamma$-ray and X-ray observations, i.e. $T_{X,0}<T_{90}$. For these cases, we take $t_{burst} = T_{90}$. In some other cases (e.g. GRBs 101225A and 050724), the emergence of the external shock afterglow component is lacking at the end of X-ray observation, so that only the lower limit of $t_{\rm burst}$ can be determined as the last XRT observation time. In some other cases (e.g. GRB 110503A), the X-ray light curve is dominated by the afterglow component from the beginning, and there is no overlap between $T_{90}$ and the XRT observation, we thus {\it exclude} them in them good sample but include them into the undetermined sample. }

The distribution of $t_{\rm burst}$ of the good sample is shown in Figure 5(a)\footnote{The distribution of the $t_{\rm burst}$ of the real time Swift GRB sample, as well as the fitting result of each individual GRB, is available online at \url{http://grbscience.com/tburst.html}.}. {The median value of $t_{\rm burst}$ of the good sample is 428 s, which is much longer than the peak of $T_{90}$ distribution in previous works \citep[e.g., about 20 s for the BATSE sample,][]{2000ApJS..126...19P}. Within the entire sample, about 25.6\% GRBs have $t_{\rm burst}>10^3$ s and 11.5\% GRBs have $t_{\rm burst}>10^4$ s.} Interestingly we found the traditional short GRBs (with T$_{\rm 90} \leq$ 2 s ) in our good sample have similar values of $t_{\rm burst}$ (blue solid line in Figure 5a) to long GRBs.

{The distribution of the $t_{\rm burst}$ of the good sample can be fitted by a mixture of two normal distributions in log space\footnote{{ We used the log-Normal function to model the $t_{\rm burst}$ components based on the 
facts that the burst duration likely depends on many physical parameters 
(e.g. mass, spinning velocity, metallicity of the progenitor star, total 
energy budget etc). Those parameters can easily play as product form 
into the function of the $t_{\rm burst}$ \citep[see e.g., ][]{2009ApJ...703.1696Z}.  Statistically speaking, if a parameter depends on the product of more than 
three random variables, then its distribution should be log-normal due to central limit theorem \citep[see e.g., ][]{Aitchison57,2002ApJ...570L..21I}.}}, with a narrow, significant peak at $\sim$ 355 s, and a wider, less significant peak at $\sim$ $2.8\times 10^4$ s respectively\footnote{We use software 'mclust', which is an R package for normal mixture modeling via expectation-maximization (EM) algorithm, to automtically indentify the optimized mixture model. The best model is selected based on the Bayesian Information Criterion (BIC). For details, see \url{http://www.stat.washington.edu/mclust/} }.} 

{As discussed above, this apparent bimodal distribution is subject to strong selection effects due to observational biases. In the following, we address two strong selection effects in turn.}

\begin{itemize}
{
\item First, there is a {\it Swift} slewing gap between $\gamma$-ray observations (i.e., T$_{90}$) and the first XRT observation time, $T_{X,0}$. It is likely that $t_{\rm burst}$ falls into this gap for many GRBs in the undetermined sample (e.g. GRB 110503A in Figure 4). The inclusion of this sample (whose size is comparable to the good sample) would modify the $t_{\rm burst}$ distribution significantly. In order to check how this effect changes the $t_{\rm burst}$ distribution we perform the following tests: 

(1) We simply let $t_{\rm burst} = {T_{90}}$ for the undetermined sample and plot the distribution of $t_{\rm burst}$ of the whole sample (good + undetermined) of 647 GRBs in Figure 5 (b). By doing so, the values of $t_{\rm burst}$ in the undetermined GRB sample could be underestimated, so that Figure 5 (b) may be still regarded as a biased illustration of the $t_{\rm burst}$ distribution. Under this treatment, these $t_{\rm burst}$ values are more consistent with a single component. However, a Gaussian model can only poorly fit the data: there appears a sudden drop of $t_{\rm burst}$ around 1000 s and a significant excess in the ``ultra-long" regime with $t_{\rm burst} \ge 10^4 $ s.

 (2) By assuming ${T_{90}} \le t_{\rm burst} \le T_{X,0}$, we generate a uniformly-distributed random value of $t_{\rm burst}$ between ${T_{90}}$ and $T_{X,0}$ in logarithmic scale and assign it to $t_{\rm burst}$ for each GRB in the undetermined sample. We then plot the the $t_{\rm burst}$ distribution of the whole sample (good + undetermined) in Figure 5 (c). A Gaussian fit is improved, but the excess of the ultra-long GRBs still exists. 

\item There is an orbital gap around thousands of seconds (Fig.4, e.g. GRB 110503A) due to various reasons such as geometry configuration between \swift orbital position relative to the GRB source position which is subject to Sun, Moon and Earth observation constraints, instrumental temperature of \swift, and delay of observation in respect to the priority of other ongoing (Target of Opportunities (ToOs). All these factors act as a selection effect against finding  $t_{\rm burst}$ values within this gap. This gap (starting from $t_{\rm gap,1}$ and ending at $t_{\rm gap,2}$, which are measured in the observed light curves, see e.g GRB 110503A in Figure 5) has a typical value of $\sim$ 3200 s (Figure 6a). The existence of such a gap has two effects on the $t_{\rm burst}$ distribution. First, if $t_{\rm burst}$ falls into this gap, these values are not registered, so that one would expect a dip in the $t_{\rm burst}$ distribution. Second, for those bursts whose real $t_{\rm burst}$ falls into this gap, one would mistakenly take an earlier steep-to-shallow transition break as $t_{\rm burst}$, giving rise to a pile-up effect before the beginning of the orbital gap (see Figure 6b), which may be responsible for the sharp drop of the $t_{\rm burst}$ distribution around 1000 s in Fig.5(b). In order to test these speculations, we perform a Monte-Carlo simulation by assuming that the intrinsic $t_{\rm burst,int}$ distribution is a single-peak Gaussian distribution in logarithmic space. Guided by the fit in Figure 5(c), we assume that the Gaussian distribution has a mean value $\mu= \log t_{\rm burst,int} = 2.2$ and a standard deviation $\sigma$=0.6. We generate $10^4$ GRBs whose $t_{\rm burst,int}$ follows such a distribution as shown in Figure 7(a). Each simulated GRB has a parameter set of \{$t_{\rm burst,int}$, $T_{90}$, $t_{gap,1}$, $t_{gap,2}$\}, where $T_{90}$, $t_{gap,1}$, $t_{gap,2}$ are generated following their corresponding observed distributions, as shown in Figure 7(b) and Figure 7(c). To take account of the orbital gap effect, we check whether each $t_{\rm burst,int}$ falls into the gap between $t_{gap,1}$ and $t_{gap,2}$ for each simulated GRB. If not, we take the ``observed'' value $t_{\rm burst} = t_{\rm burst,int}$. If yes, we then assign $t_{\rm burst}$ a random value between $T_{90}$ and $t_{gap, 1}$ in the logarithmic scale. The distribution of the final simulated $t_{\rm burst}$ is shown as the solid line in Figure 7(d), where the intrinsic input distribution is also plotted as the red dotted histogram. The resulting simulated the $t_{\rm burst}$ distribution shows a significantly sharp drop around 1000-3000 s as well as dip afterwards. All these signatures are similar to the $t_{\rm burst}$ distributions derived from the data (Fig.5(a-c)). Our simulation suggests that the hypothesis of one single $t_{\rm burst}$ distribution component cannot be ruled out by the data.
}
\end{itemize}

\section{Summary and Theoretical Implications}
{In this paper, we investigate the true GRB central engine activity duration distribution by considering both $\gamma$-ray and X-ray data. By defining $t_{\rm burst}$ based on some physically motivated criteria, we robustly derived $t_{\rm burst}$ for 343 GRBs. The $t_{\rm burst}$ distribution of this ``good'' sample shows an apparent bimodal distribution. If this is true, ultra-long GRBs could be more common than suggested in the literature (e.g., Levan et al. 2014). However, by including a larger sample whose $t_{\rm burst}$ values are not measured but can be guessed (303 GRBs in the ``undetermined'' sample) and by addressing two important selection effects, we found that the intrinsic $t_{\rm burst}$ distribution can be consistent with one single component. The existence of a separate ``ultra-long'' category of GRBs \citep{Levan:2013vr,Gendre:2013tr,2013arXiv1310.4944B}is neither required nor excluded by the data. Our results suggest that 
the ultra-long GRBs could be just a tail of a single long-duration GRB sample \citep[see also][]{2013arXiv1310.0313V}. }

As shown in Figure 8, our result indicates that a large fraction of long GRBs are actually quite long, even though their $T_{90}$'s are not extremely long. Evidence that two such long GRBs (030329 and 130427A) have associated Type Ic supernovae \citep{2003ApJ...591L..17S,Hjorth:2003jv,2013ApJ...776...98X} suggest that their progenitor is likely a Wolf-Rayet star whose hydrogen and helium envelopes have been depleted.
The fact that their $T_{90}$'s are much longer than 10 s, the typical time scale for the jet to penetrate through the stellar envelope, suggests that the burst duration is not necessarily related to the size of the progenitor. Hence, making a direct connection between ultra-long GRBs and blue supergiants progenitor lacks strong physical justification.
Theoretical investigations show that it becomes much more difficult for a jet to successfully penetrate through the stellar envelope of a blue supergiant, so that a significant fraction of such collapsing stars may just lead to failed GRBs \citep{Murase:2013fo}. Also, blue supergiants are very unstable and short-lived, and their final explosion properties, including the possibility of launching a jet remain unclear.

How to prolong a GRB central engine duration with a compact progenitor star is an open question. For variable emission such as X-ray flares, fragmentation in the massive star envelope \citep{King:2005il}, fragmentation in the accretion disk \citep{2006ApJ...636L..29P}, and the formation of a magnetic barrier around the accretor \citep{2006MNRAS.370L..61P} have been proposed. If the engine is a millisecond magnetar instead of a black hole, the magnetic activity of the millisecond magnetar can power an extended emission \citep{2011MNRAS.413.2031M}. The steady spin down of the magnetar \citep{2001ApJ...552L..35Z} would also power an internal X-ray plateau \citep{2007ApJ...665..599T}. Alternatively, fall-back accretion of the stellar envelope onto a newly formed black hole \citep{Kumar:2008fa,2041-8205-767-2-L36} can also make extended internal X-ray emission. All these mechanisms could also be applied to ultra-long GRBs without invoking a large progenitor star.

The wide peak of ultra-long GRB components may be also understood in a scenario where those GRB progenitor stars have a distribution of mass and size, ranging from Wolf-Rayet stars to blue supergiants.. Further multi-wavelength data, especially the properties of associated SNe and host galaxies of GRBs with different $t_{\rm burst}$, are needed to make further progress.

\cite{Bromberg:2013ia} found a plateau in the $dN/d T_{90}$ distribution in the BATSE, \swift and {\it Fermi} GBM samples, and argued that it provides direct evidence of the collapsar model. Realizing that $T_{90}$ is no longer a good indication of central engine activity time scale, we apply our $t_{\rm burst}$ data {\rm in the good sample} to carry out a $dN/d t_{\rm burst}$ analysis. The plateau found by Bromberg et al. using T90 is not reproduced with $t_{\rm burst}$ (Figure 9). Admittedly, the jet power in most GRBs reduces with time, and the most energy is still released during $T_{90}$. In any case, the collapsar signature suggested by \cite{Bromberg:2013ia} may need further investigation.

\acknowledgments

We thank an anonymous referee for thoughtful comments, and David N. Burrows, Peter M\'esz\'aros, Xiao-Hong Zhao, Peter Veres, Kazumi Kashiyama, Xue-Wen Liu, Derek Fox and Shaolin Xiong for helpful discussion and suggestions. We thank Dirk Grupe for the information about \swift operations. BBZ thanks Jason Rudy for helpful comments on the codes of multivariate adaptive regression splines fitting. This work was partially supported by NASA /Fermi GI grant/ NNX11AO19G (BBZ). B.Z. acknowledges support from NASA NNX10AP53G. KM acknowledges the support by NASA through a Hubble Fellowship, Grant No. 51310.01 awarded by the STScI, which is operated by the Association of
Universities for Research in Astronomy, Inc., for NASA, under Contract
No. NAS 5-26555. We acknowledge the use of public data from the \swift data archive.
{\it Facilities:} \facility{\swift}.

\begin{figure}[ht!]

\begin{center}

\includegraphics[scale=0.40]{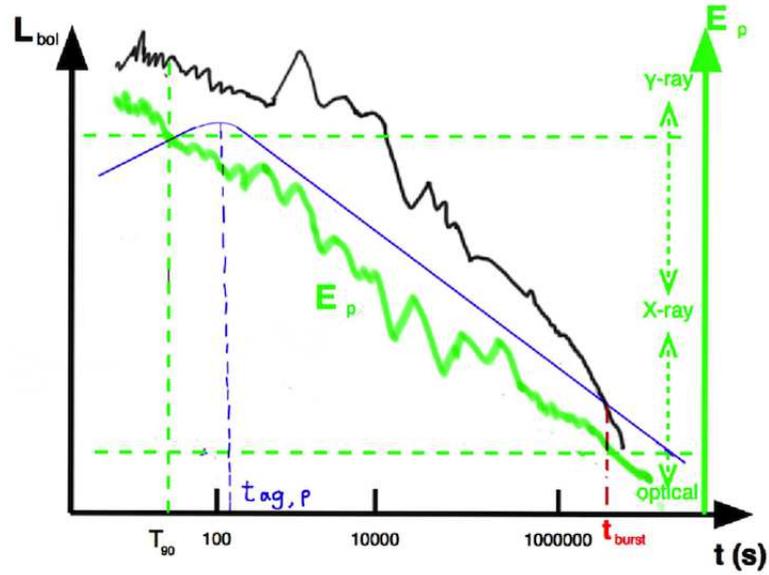}

\end{center}
 \caption{%
A sketch of the physical picture of GRB internal and external emission. The black curve denotes the bolometric internal emission light curve. The green solid curve denotes $E_p$ evolution of the internal emission, indicating that the internal emission is initially in the $\gamma$-ray band, but shifts to X-rays later. The blue curve represents the external-shock afterglow emission component, which peaks at $t_{ag,p}$ and becomes dominant at $t>t_{\rm burst}$.
}
\label{fig:eps_e}
\end{figure}

\begin{figure}[ht!]

\begin{center}

\includegraphics[scale=0.5]{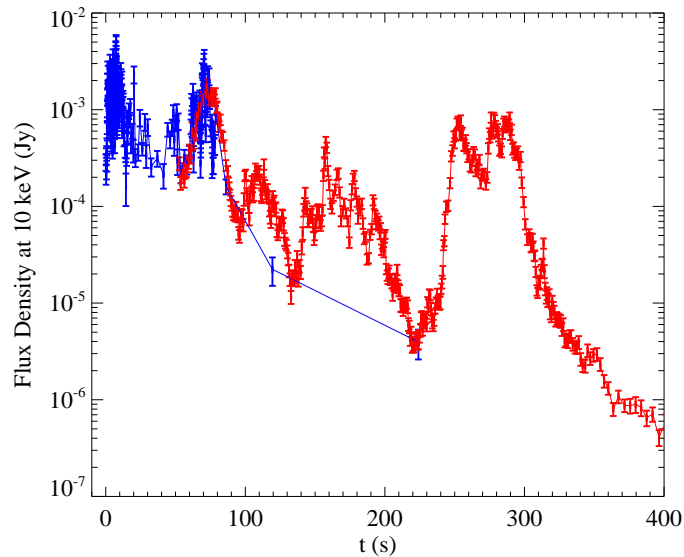}

\end{center}
 \caption{%
An example (GRB 090715B) that shows the similarity of X-ray flares (red data points) and prompt $\gamma$-ray emission (blue data points). The $T_{90}$ of this GRB is 266 s, while $t_{\rm burst}$, determined by X-ray data, is $373\pm 3$ s. Data are taken from http://www.swift.ac.uk/burst\_analyser/00357512/, where the BAT and XRT data are extrapolated to the common energy band (10 keV) using their spectral information, respectively.}
\label{fig:eps_e}
\end{figure}

\begin{figure}[ht!]

\begin{center}

\includegraphics[scale=0.55]{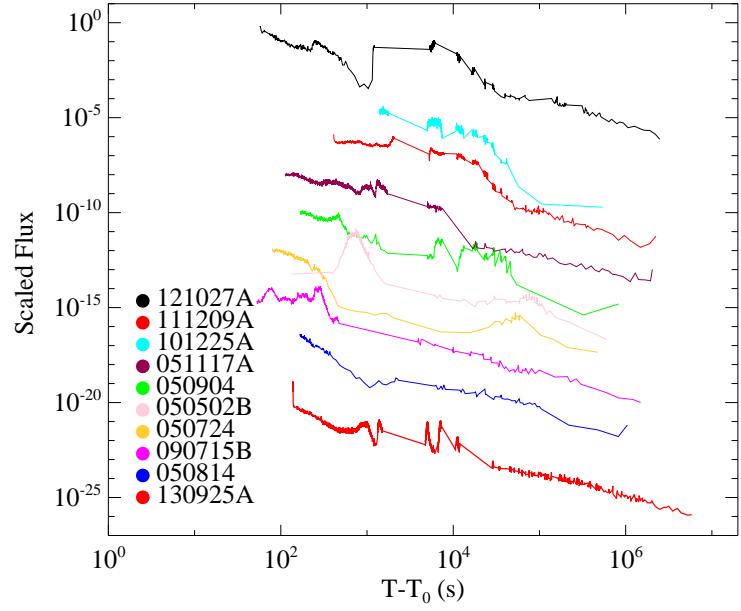}

\end{center}
 \caption{%
A comparison of $\gamma$/X-ray emission light curves of some GRBs, including the claimed
four ultra-long GRBs and some others. Two other GRBs (050904 and 051117A) also show very
similar features as the four events, suggesting that the so-called ``ultra-long'' GRBs
may not be rare events. They are likely the extreme cases of normal GRBs with bright
extended central engine activity emission.
}
\label{fig:eps_e}
\end{figure}

\begin{figure}[ht!]

\begin{center}

\begin{tabular}{ccc}

050724& 051117A & 090715B \\
 \includegraphics[scale=0.25]{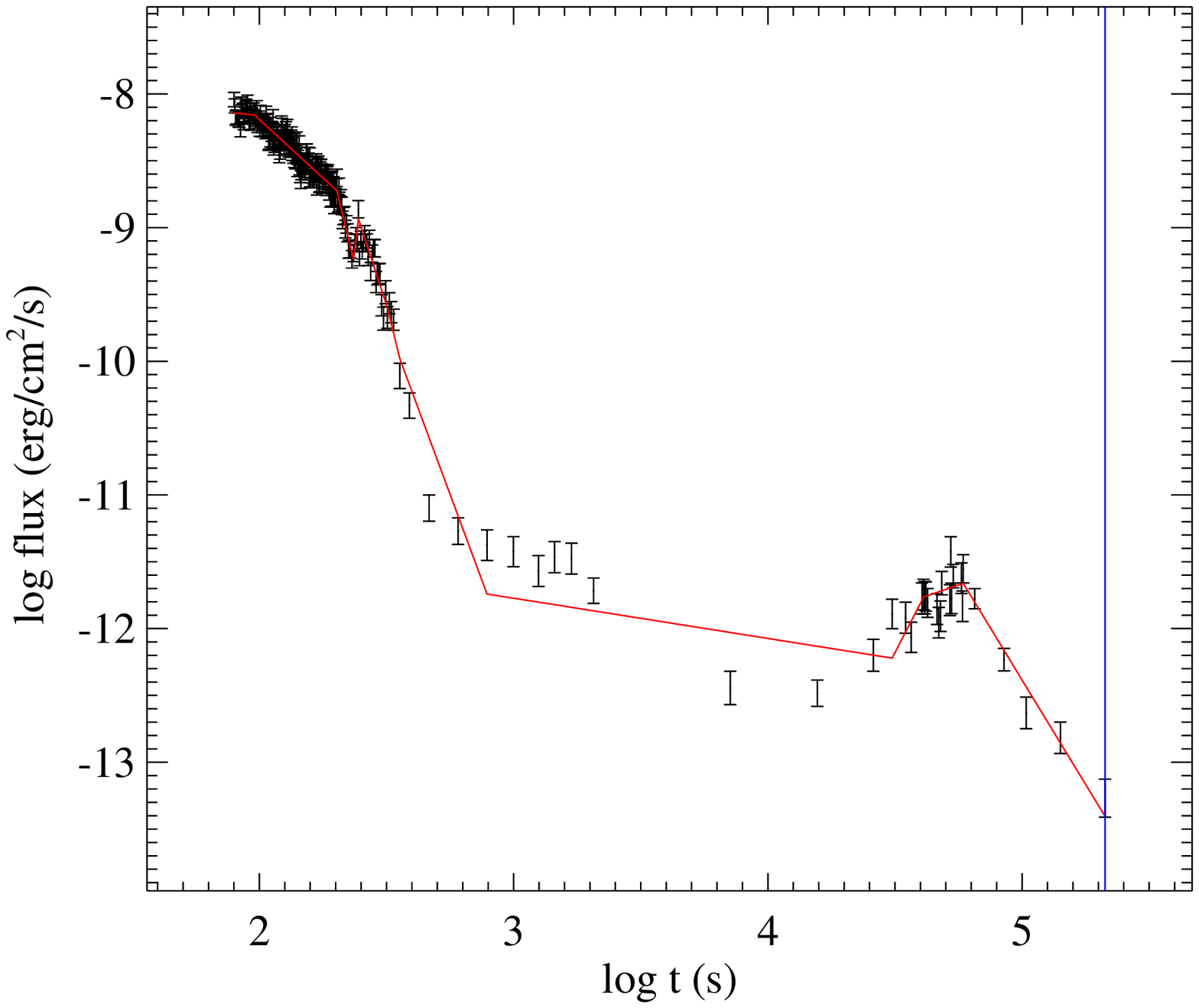} & \includegraphics[scale=0.25]{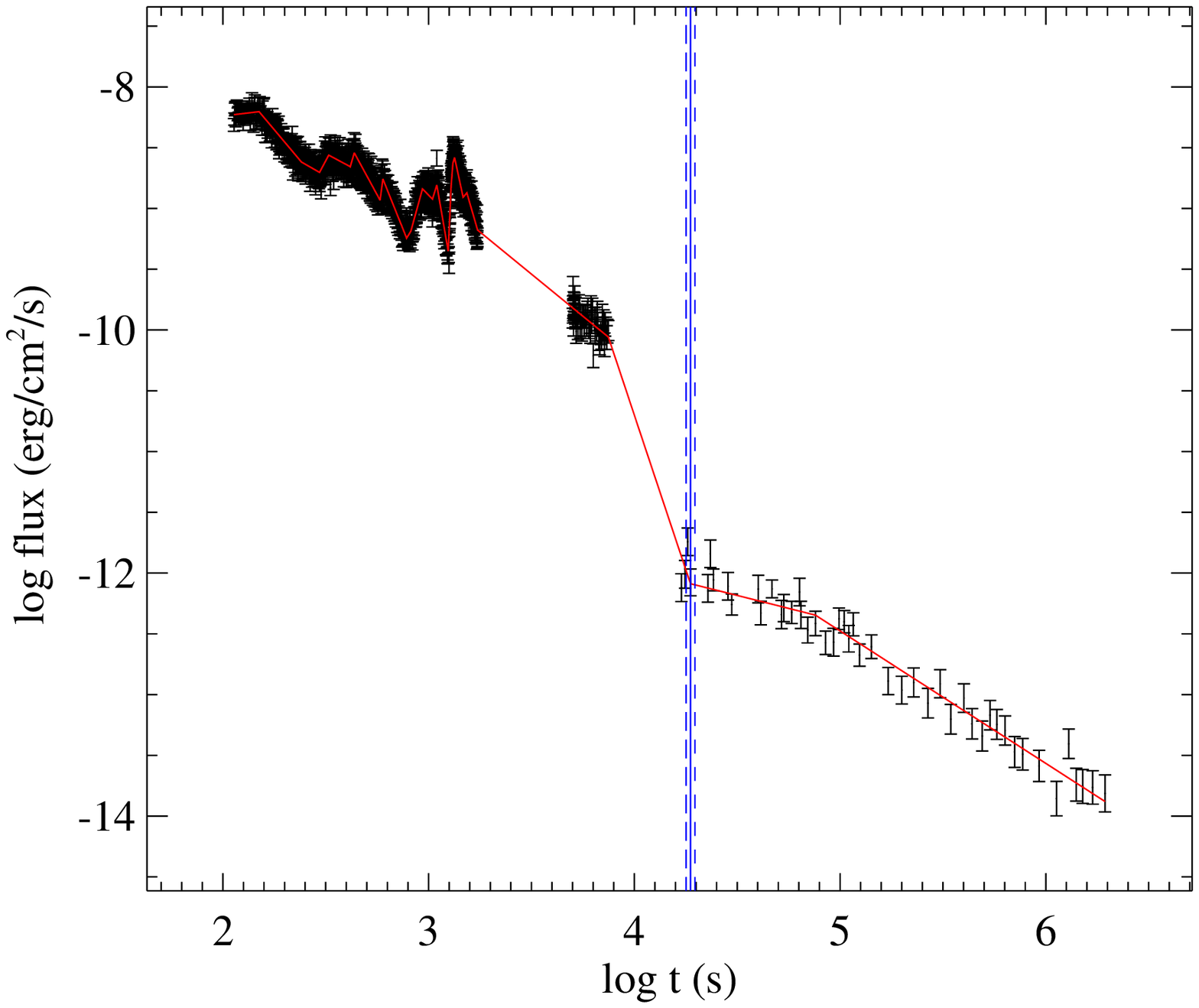} & \includegraphics[scale=0.25]{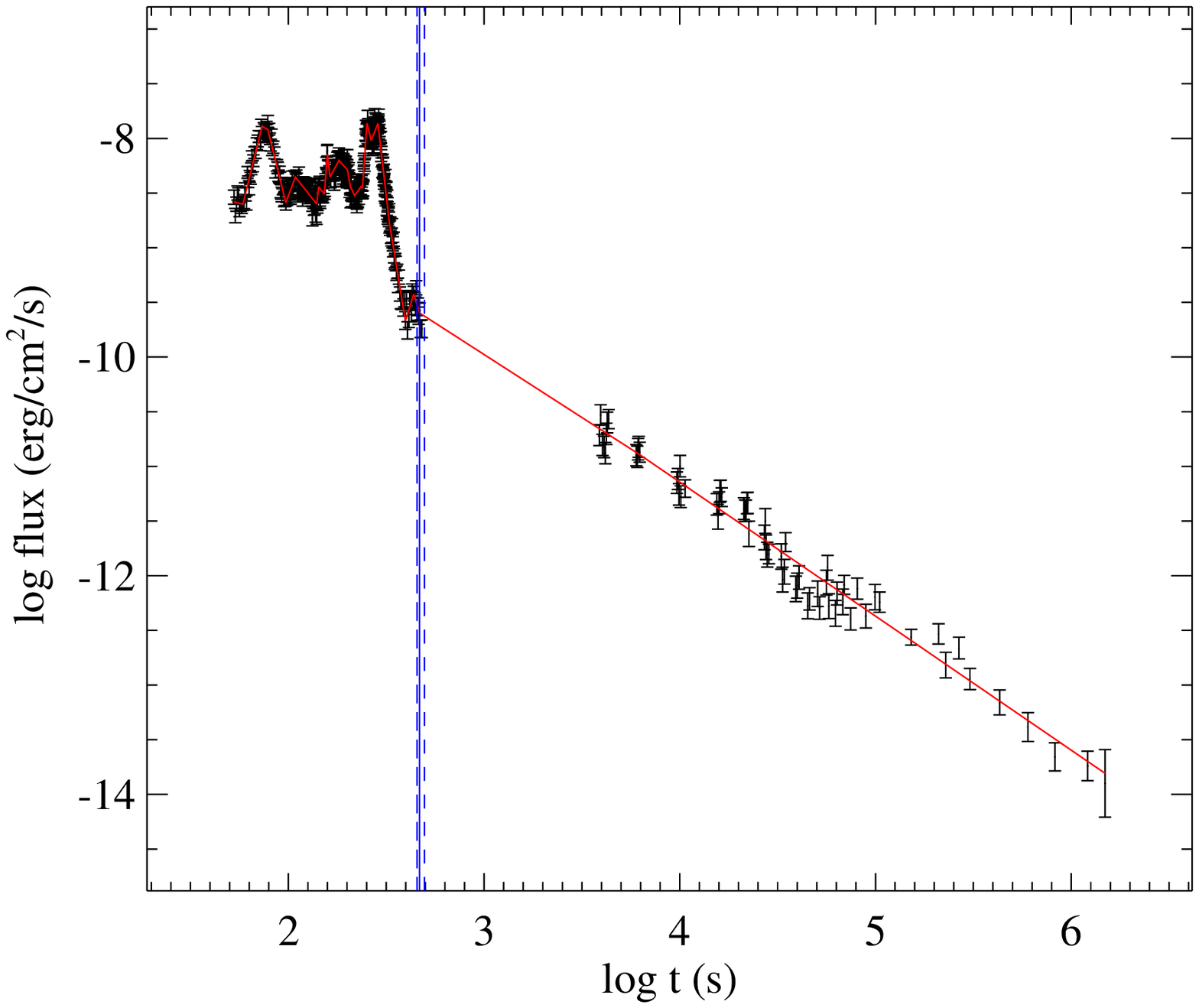} \\
101225A & 110503A & 111209A \\
\includegraphics[scale=0.25]{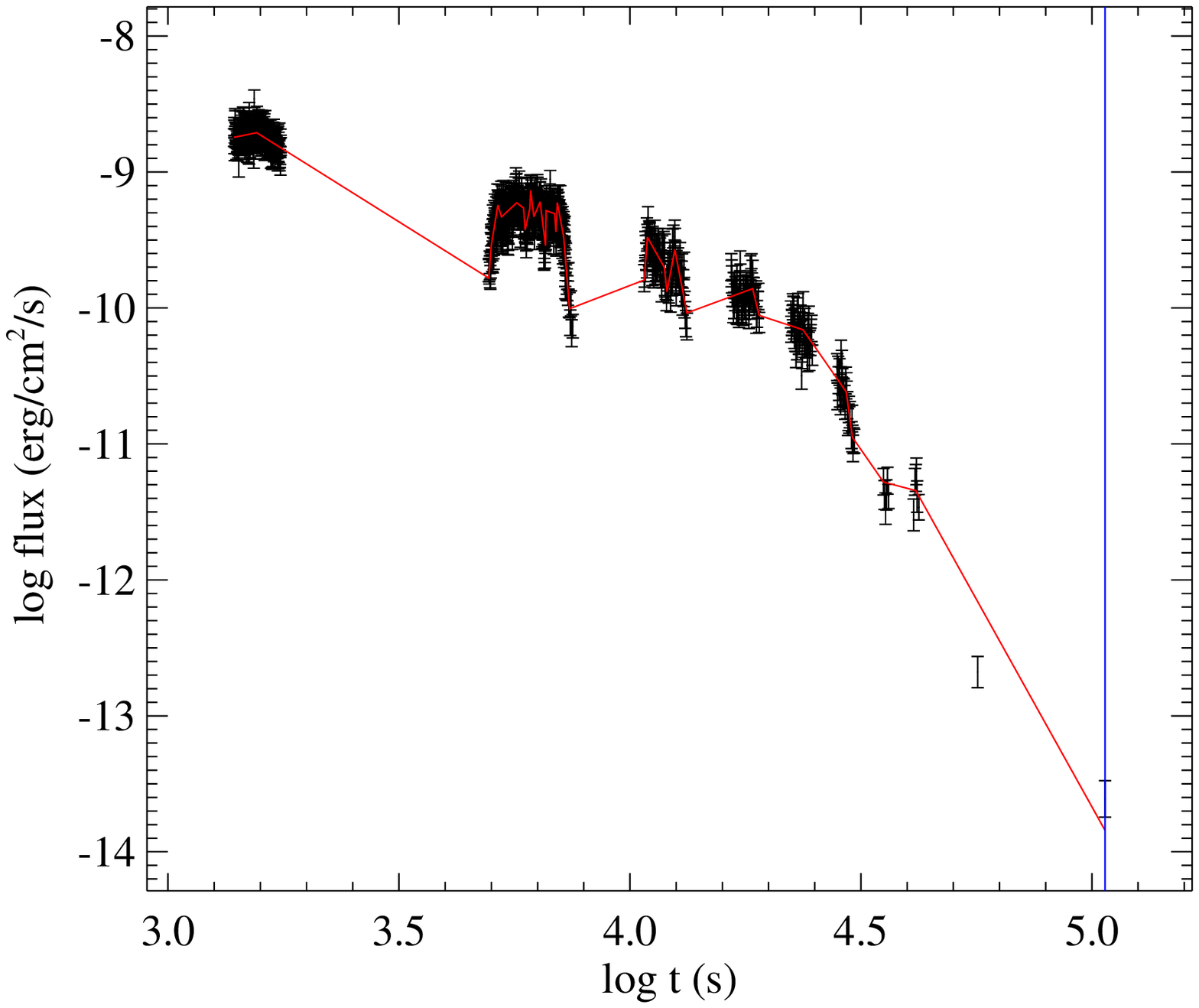} & \includegraphics[scale=0.25]{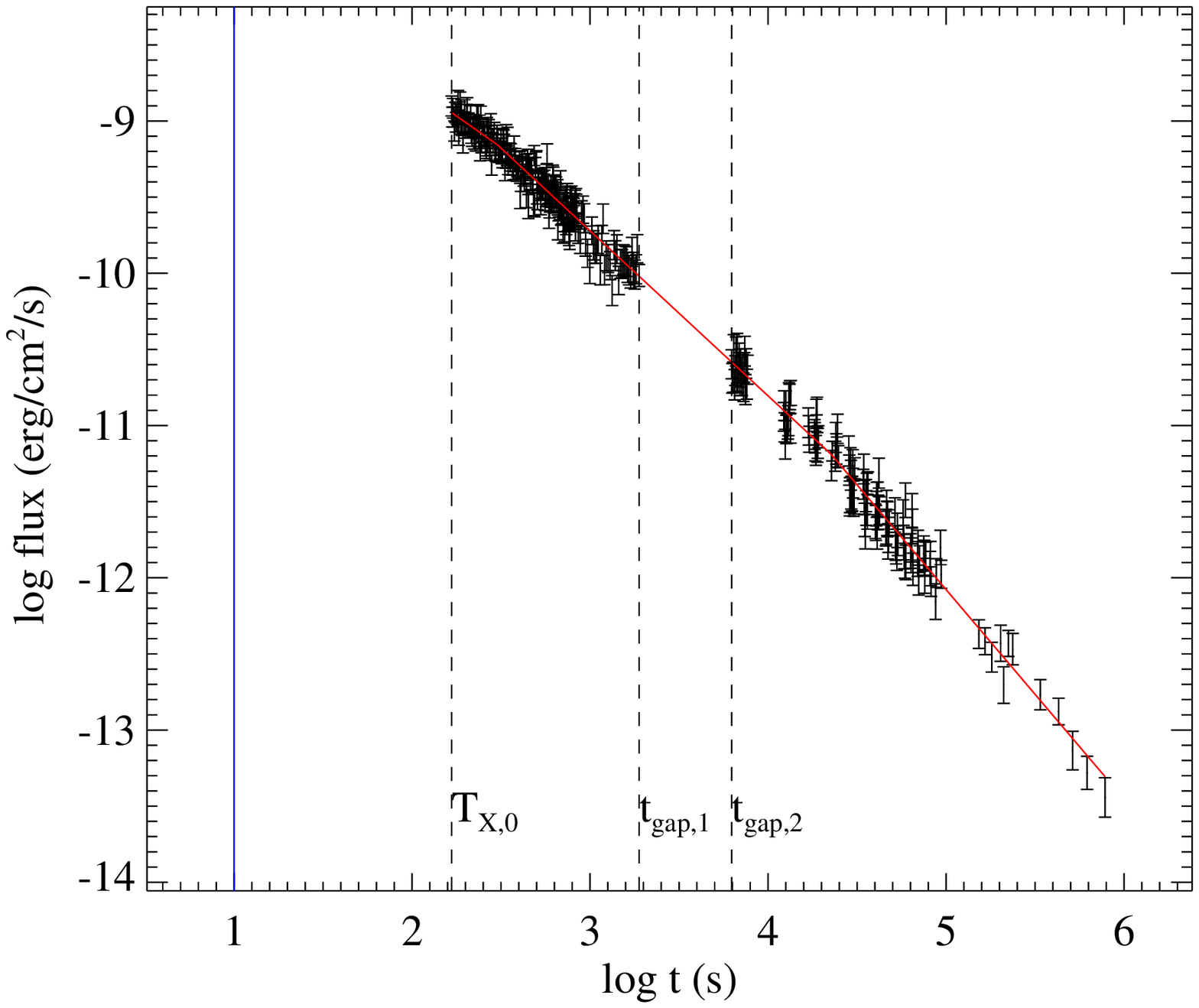} & \includegraphics[scale=0.25]{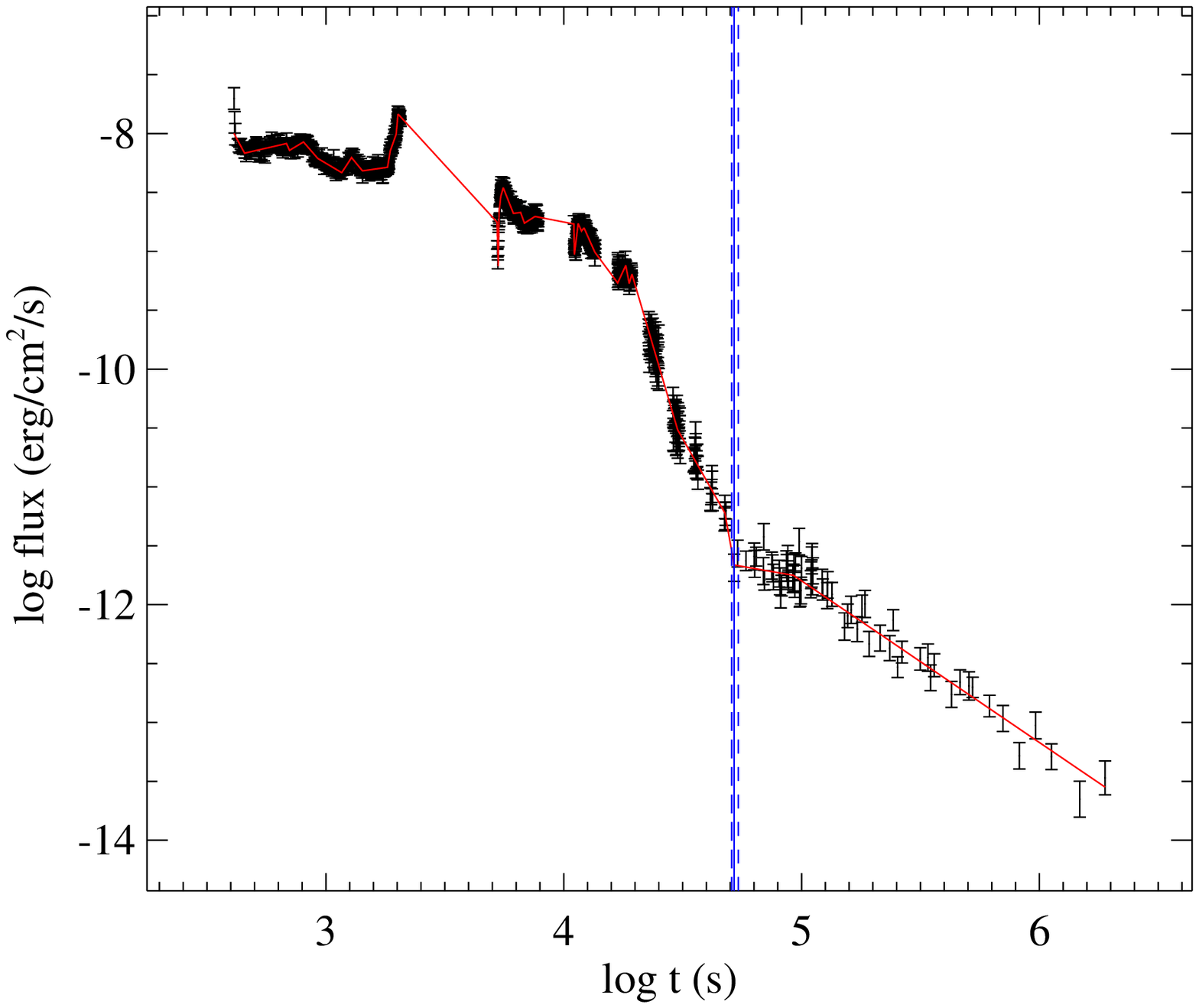} \\
121027A & 130925A & 140102A \\
\includegraphics[scale=0.25]{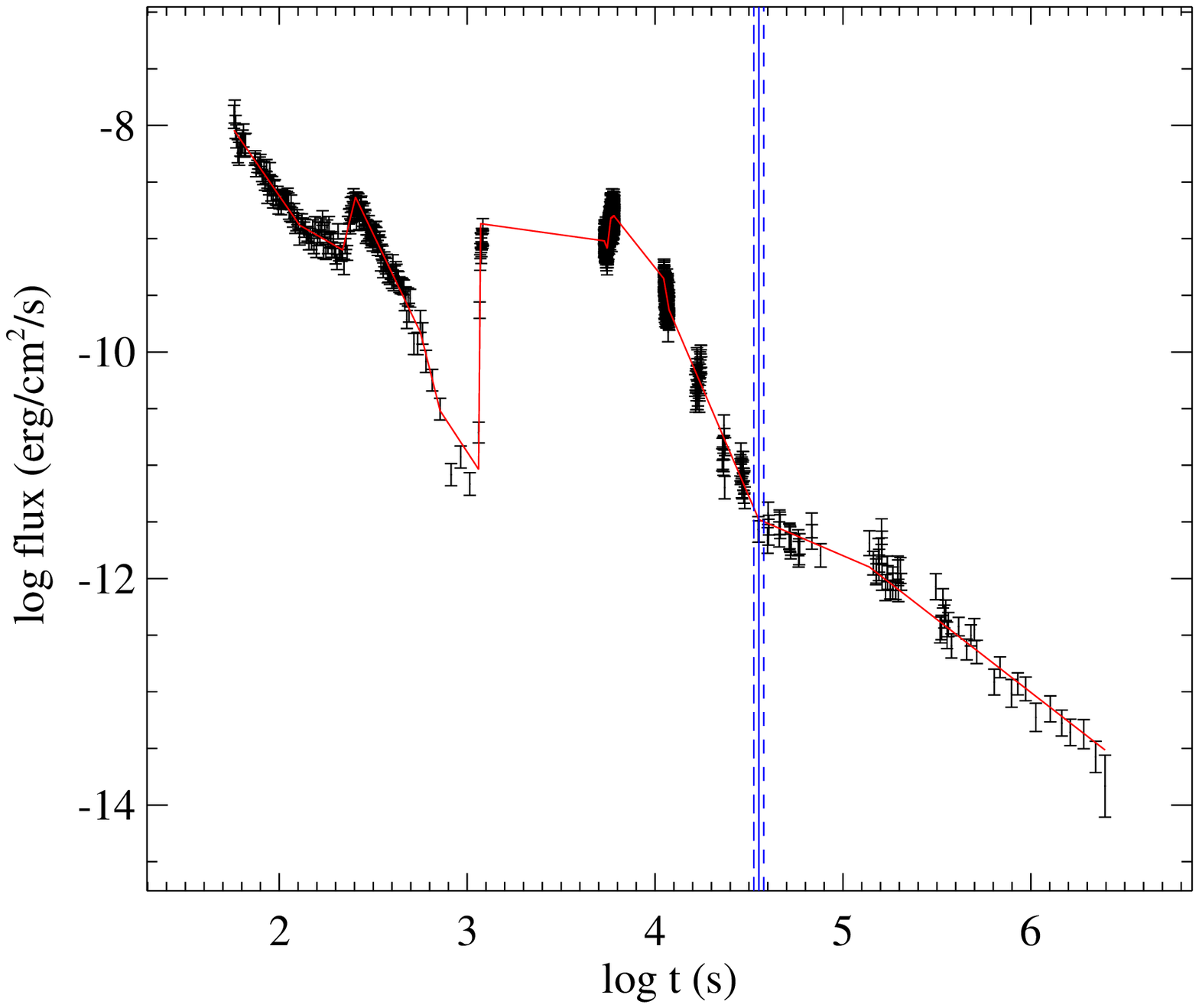} & \includegraphics[scale=0.25]{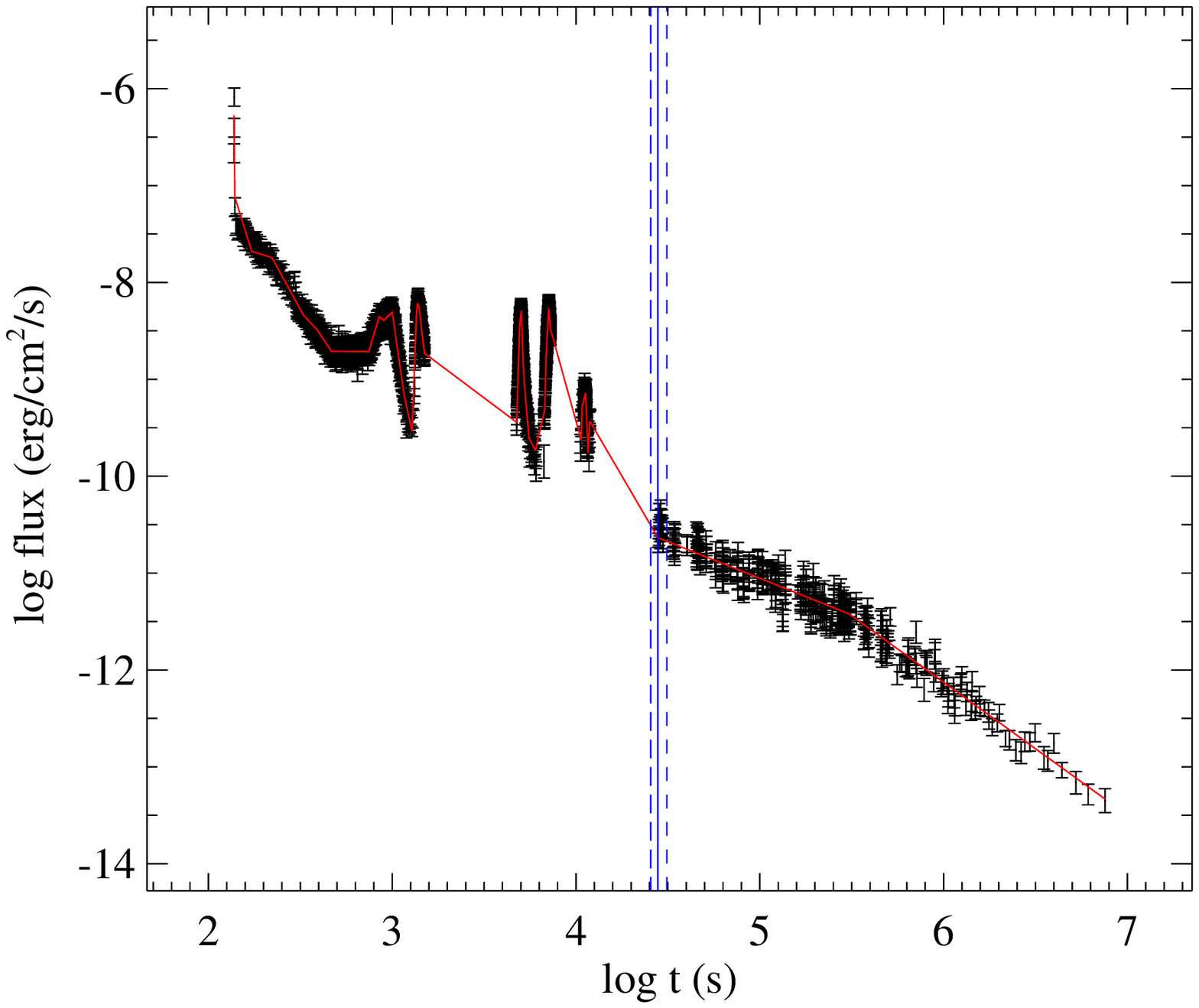} & \includegraphics[scale=0.25]{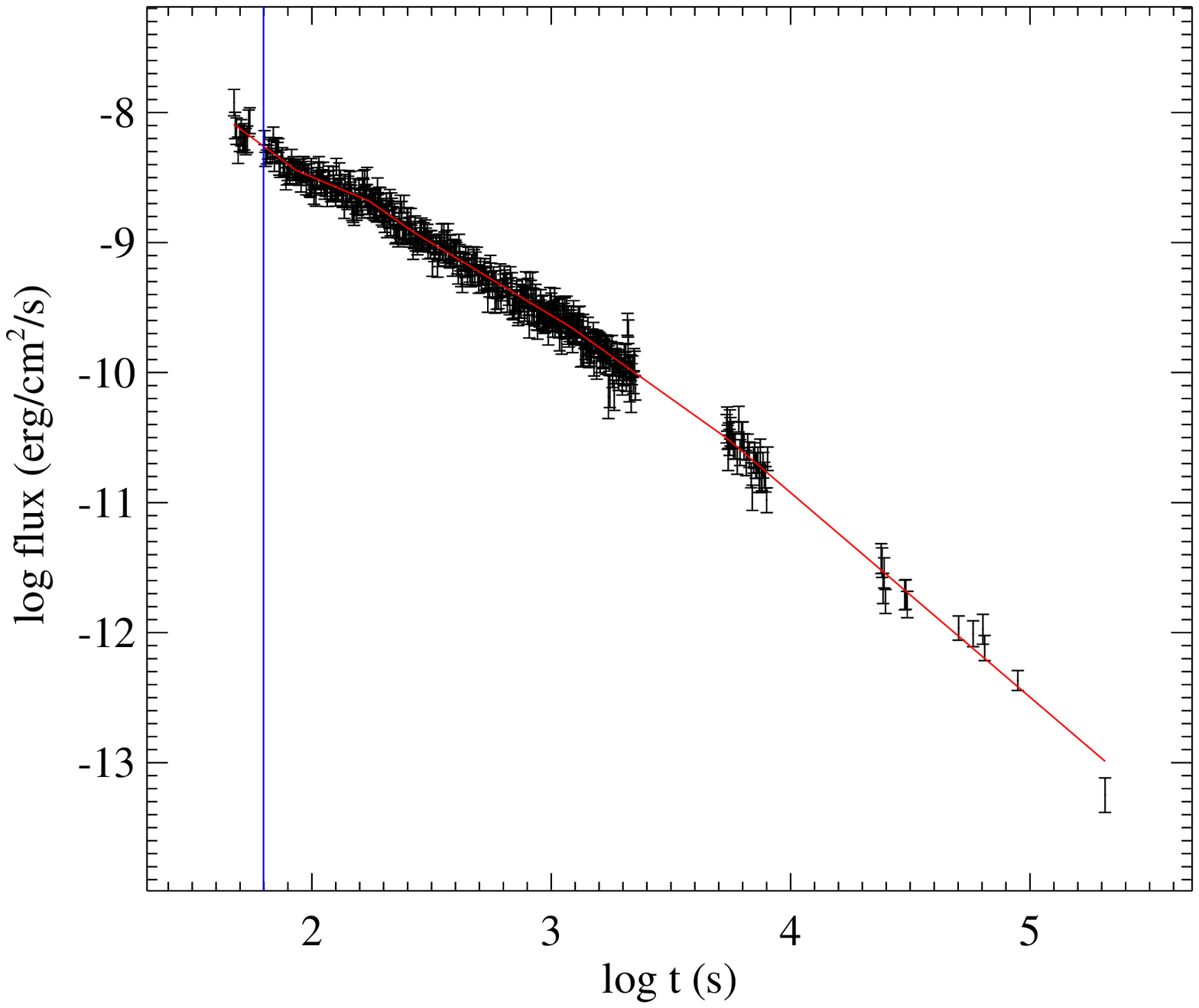}\\

\end{tabular}
\end{center}
 \caption{%
Some examples to show how to constrain $t_{\rm burst}$ with the XRT data. Black points are \swift/XRT observations. Red solid line represents the multi-segment
broken power-law model fitted to the data. Blue solid line indicates the location of $t_{\rm burst}$, and blue dashed lines (if available) represent the $1\sigma$ uncertainty of $t_{\rm burst}$. {Note that GRB 110503A is not included in the good sample but is in the undetermined sample; see \S4 for details. Data (0.3-10 keV energy flux) are taken from http://www.swift.ac.uk/xrt\_curves/.} }
\label{fig:eps_e}
\end{figure}

\begin{figure}[ht!]

\begin{center}
\begin{tabular} {ccc}
\includegraphics[scale=0.31]{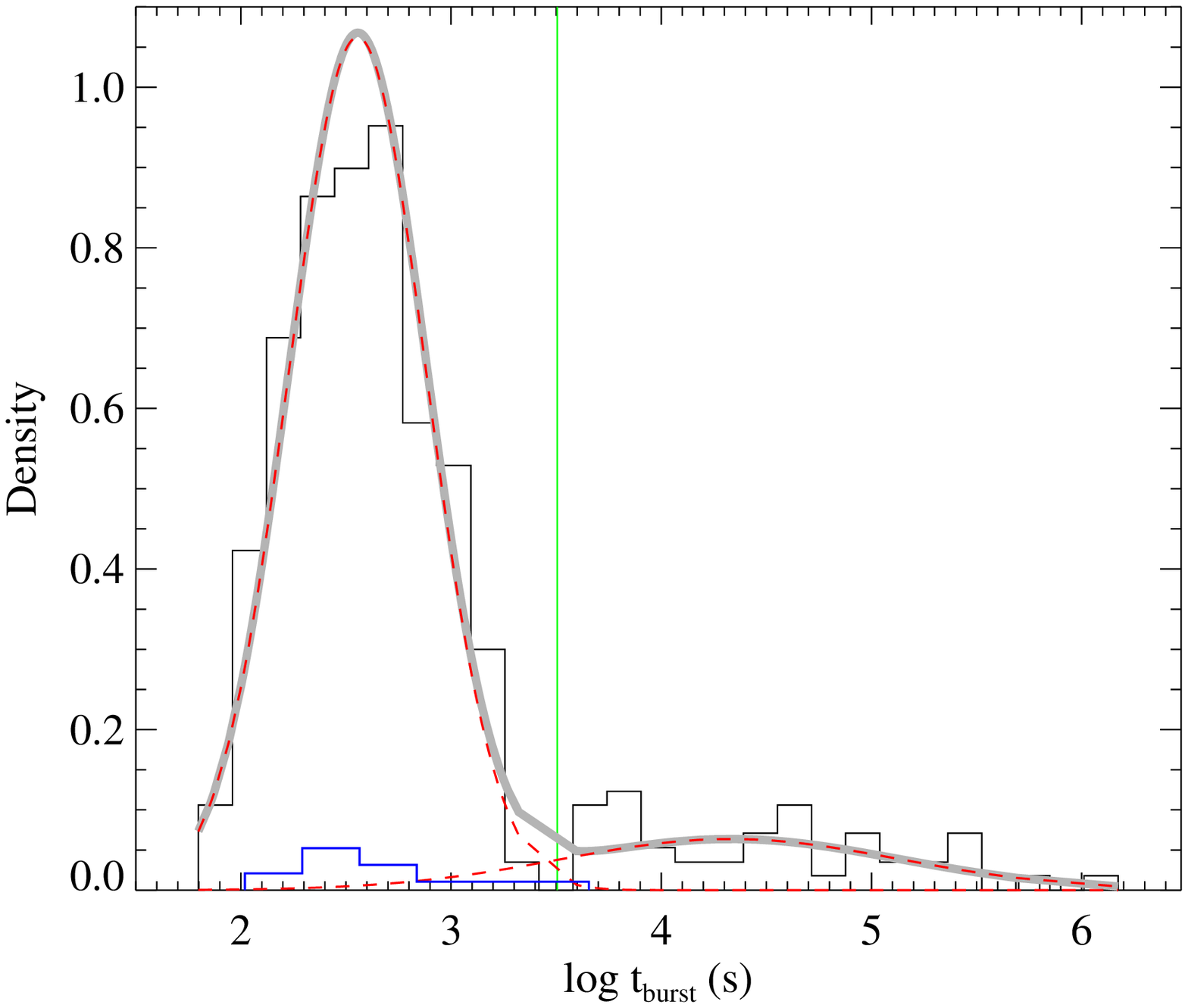} &\includegraphics[scale=0.33]{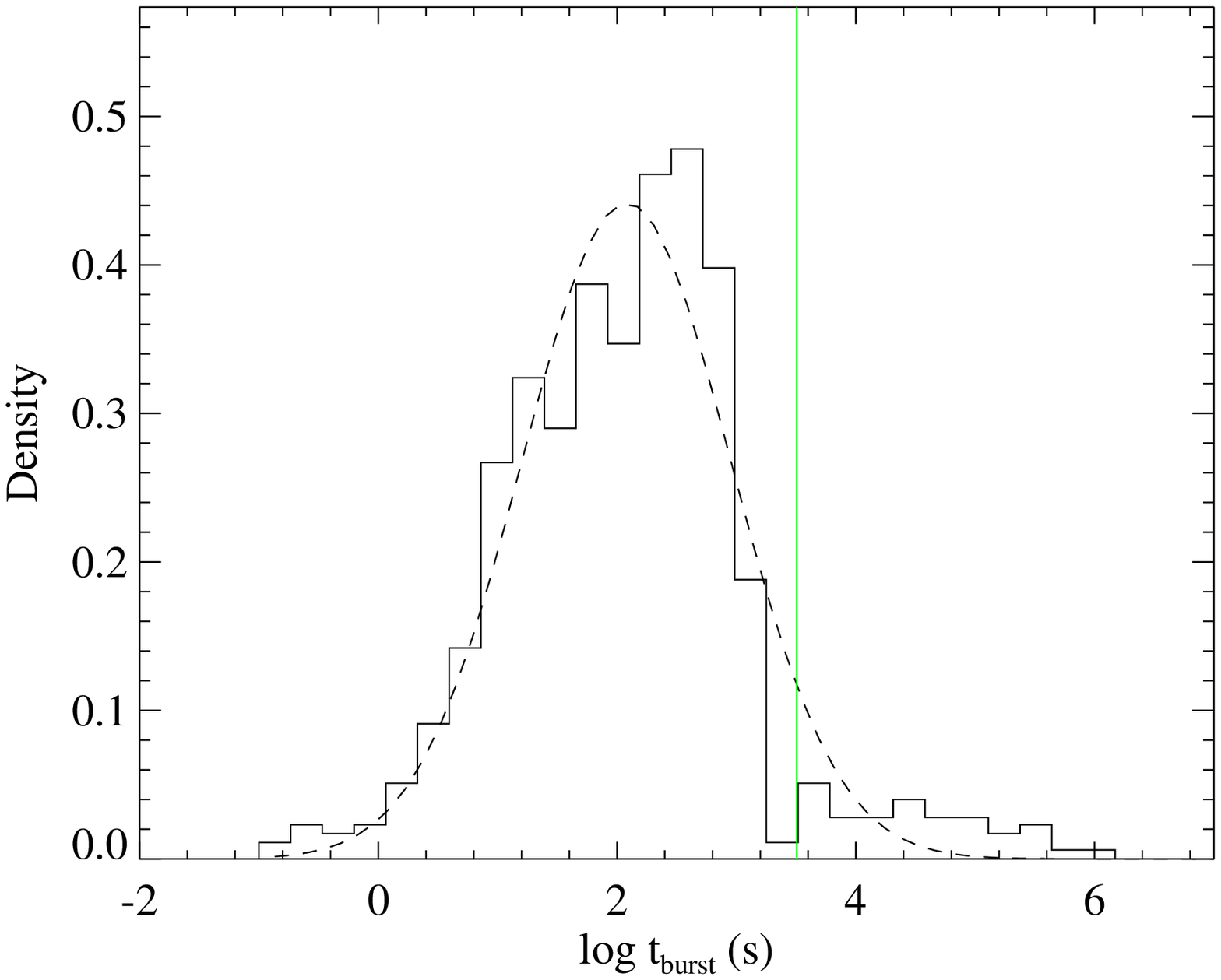}&
\includegraphics[scale=0.33]{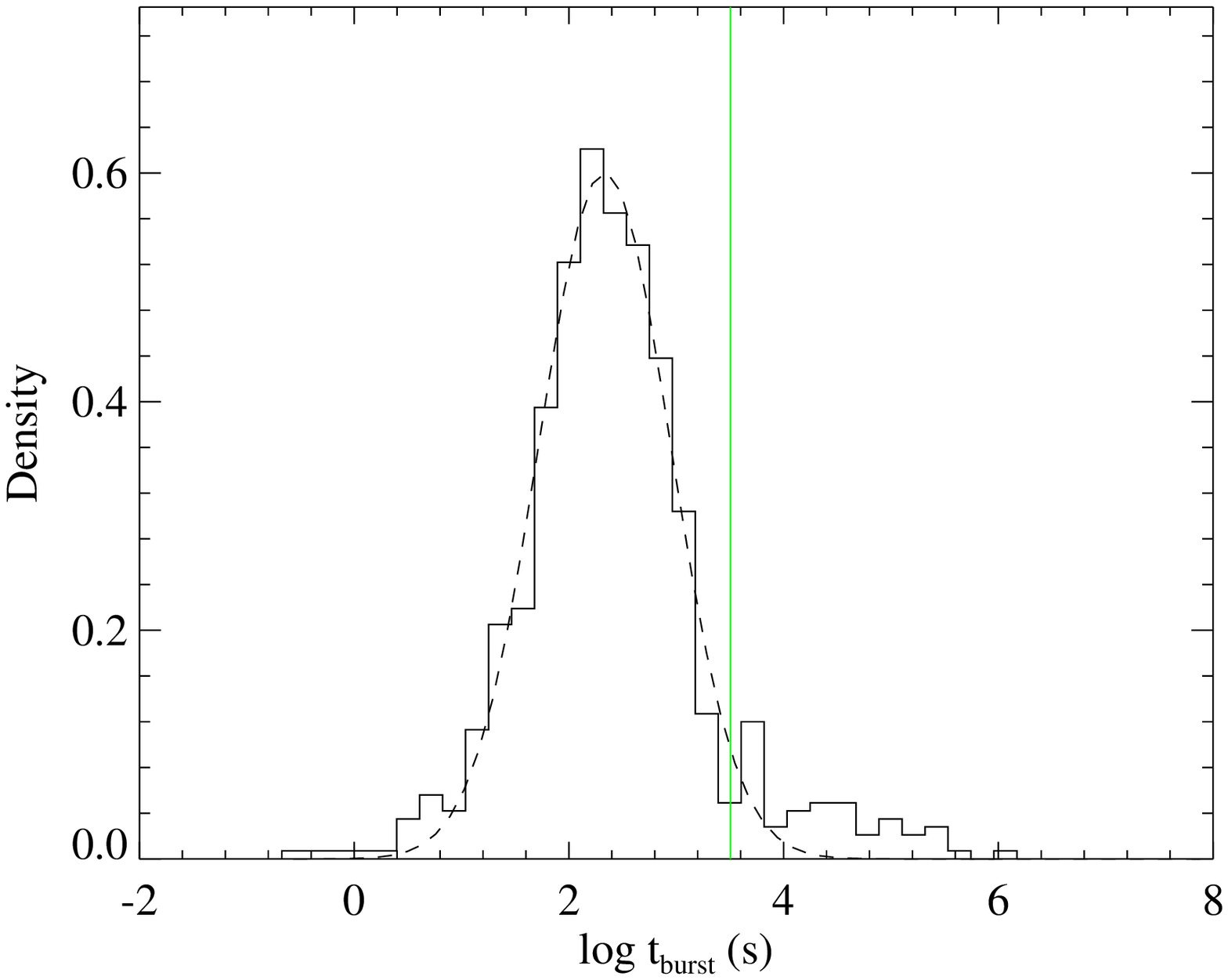} \\
(a) & (b) & (c) \\
\end{tabular}

\end{center}
 \caption{
(a) The derived distribution of $t_{\rm burst}$ of the good sample (343 GRBs). The histogram bin sizes are optimized using Knuth's rule \citep{2000cs.......11047K}. The vertical axis ``density'' is defined as ``count/bin size/total count''.
The derived $t_{\rm burst}$ are plotted as a black solid histogram. The distribution of the short GRBs (T$_{90} < $2s) in the good sample is plotted as the blue solid histogram. The fit result by a two-component Gaussian distribution is plotted as a thick grey solid line and each component is plotted as red dashed lines. {A typical value of $t_{gap,2}$ - $t_{\rm gap,1}$ = 3200 s is plotted as a vertical green solid line.} (b) Distribution of $t_{\rm burst}$ for the good sample (343 GRBs) and the uncertain sample (304 GRBs), with $t_{\rm burst}$ of the uncertain sample set to $T_{90}$. (c) Same as (b), but with $t_{\rm burst}$ in the uncertain sample set to a uniformly-distributed random value between T${_{\rm 90}}$ and T$_{\rm X,0}$ in logarithmic scale.
 }
\label{fig:eps_e}
\end{figure}

\begin{figure}[ht!]

\begin{center}
\begin{tabular} {cc}
\includegraphics[scale=0.41]{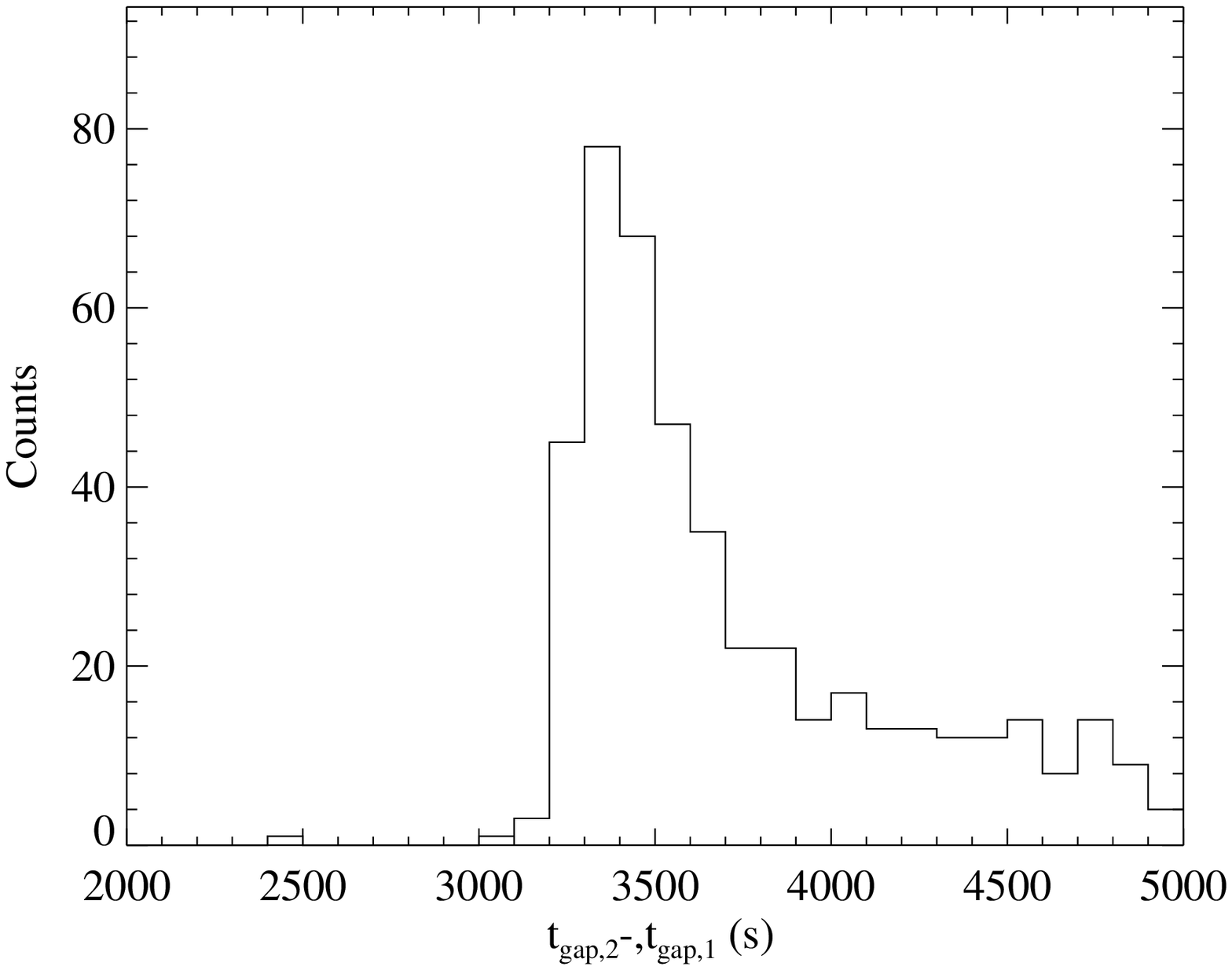} &\includegraphics[scale=0.4]{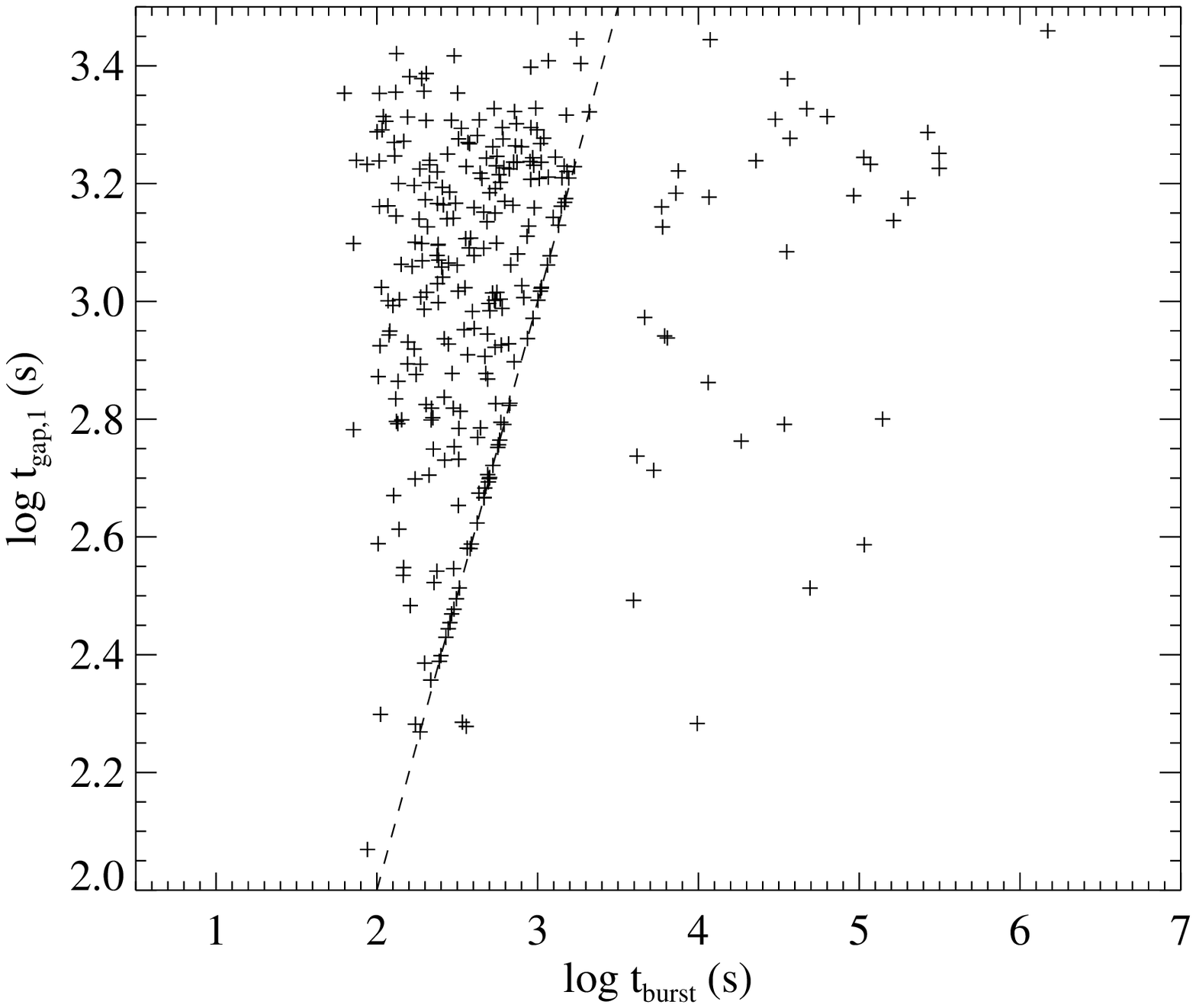}\\
(a) & (b)\\

\\
\end{tabular}

\end{center}
 \caption{
(a) Distrbution of gap times in the XRT observations of the bursts in our sample  $t_{gap,2}$ - $t_{\rm gap,1}$. $t_{gap,1}$ is the start of the gap, $t_{\rm gap,2}$ is the end; (b) comparison between $t_{\rm burst}$ between $t_{gap,1}$, which shows most $t_{\rm burst}$ are measured before $t_{gap,1}$.
 }
\label{fig:eps_e}
\end{figure}

\begin{figure}[ht!]

\begin{center}
\begin{tabular} {cc}
\includegraphics[scale=0.40]{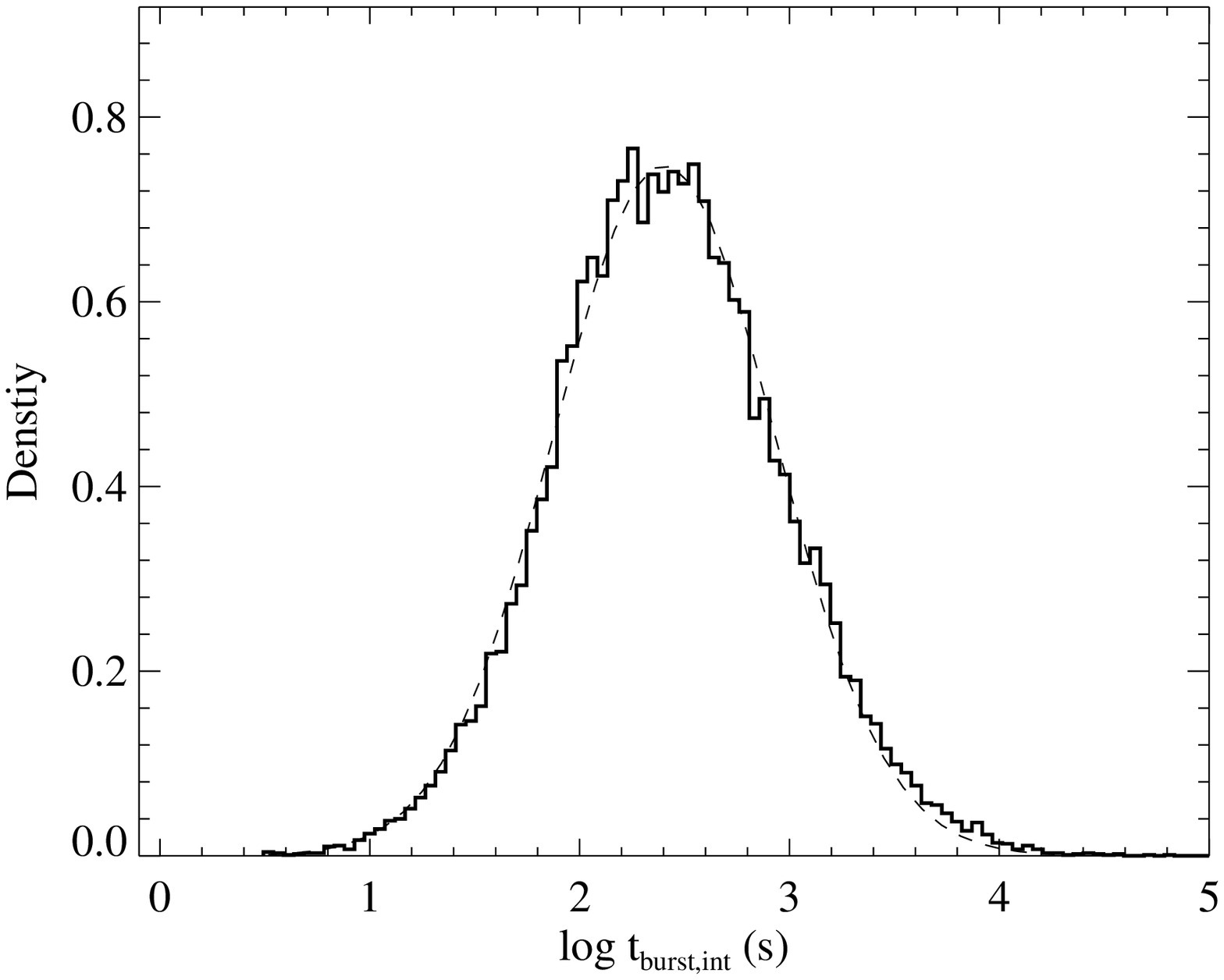} &\includegraphics[scale=0.4]{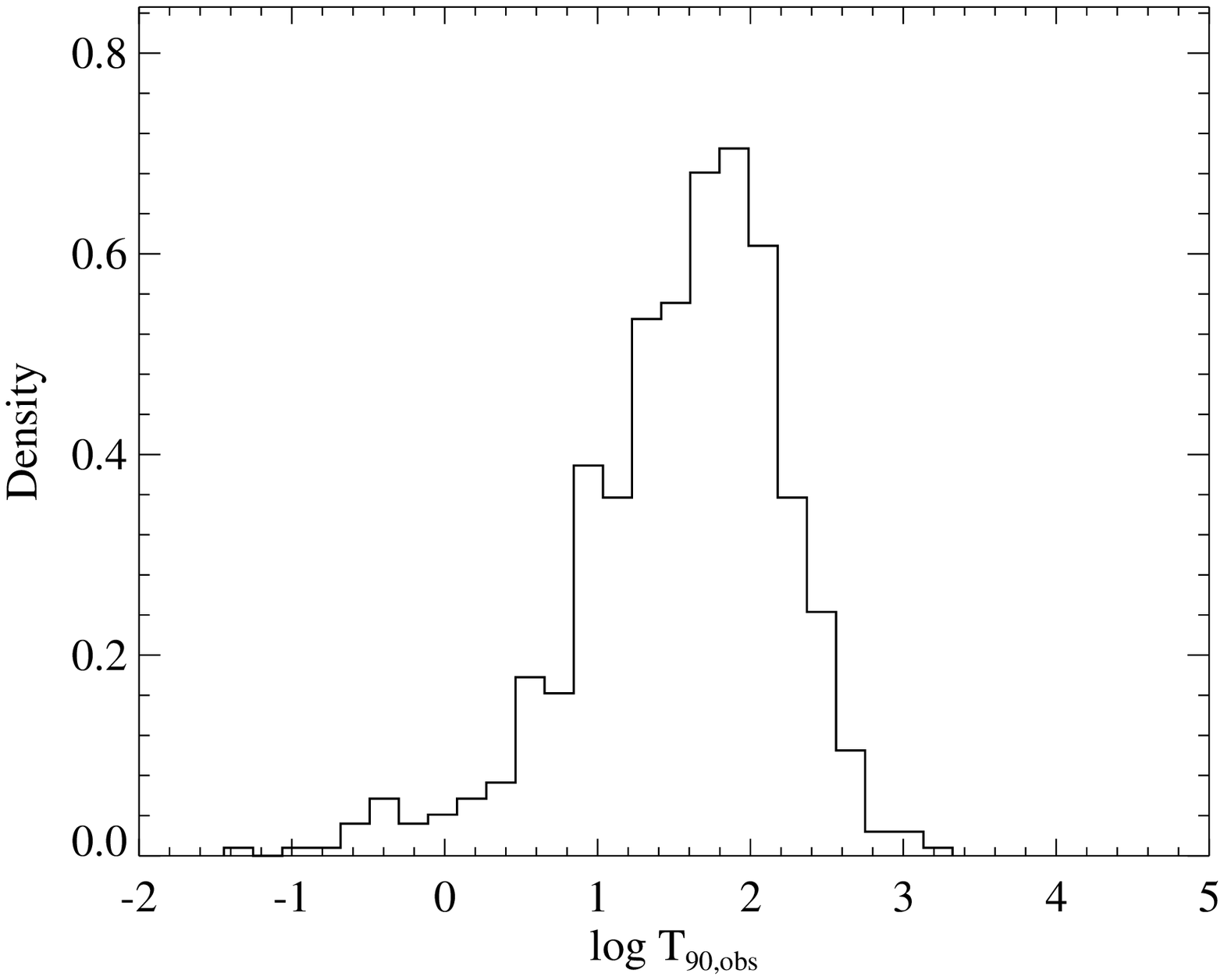}\\
(a) & (b)\\
\includegraphics[scale=0.4]{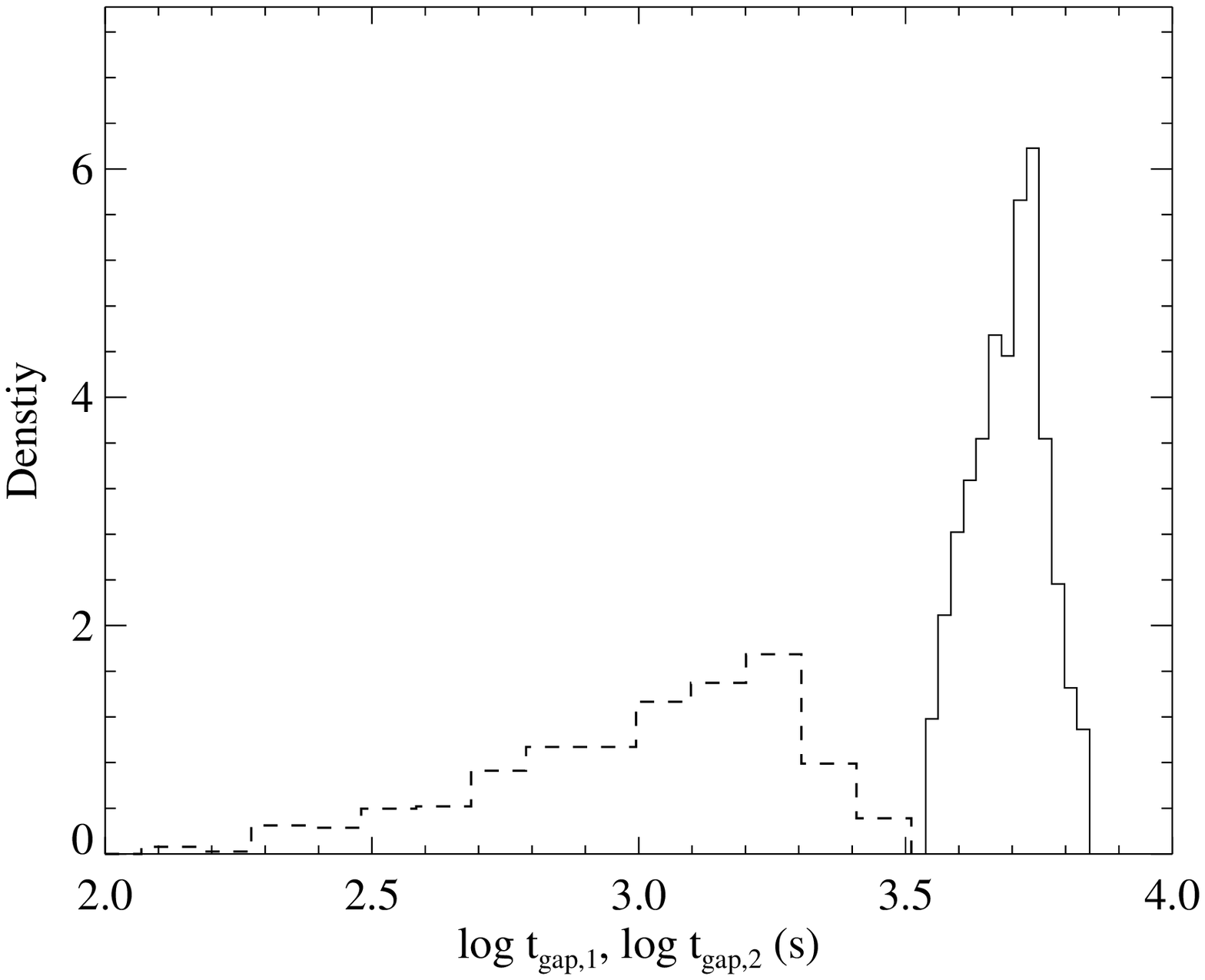} & \includegraphics[scale=0.4]{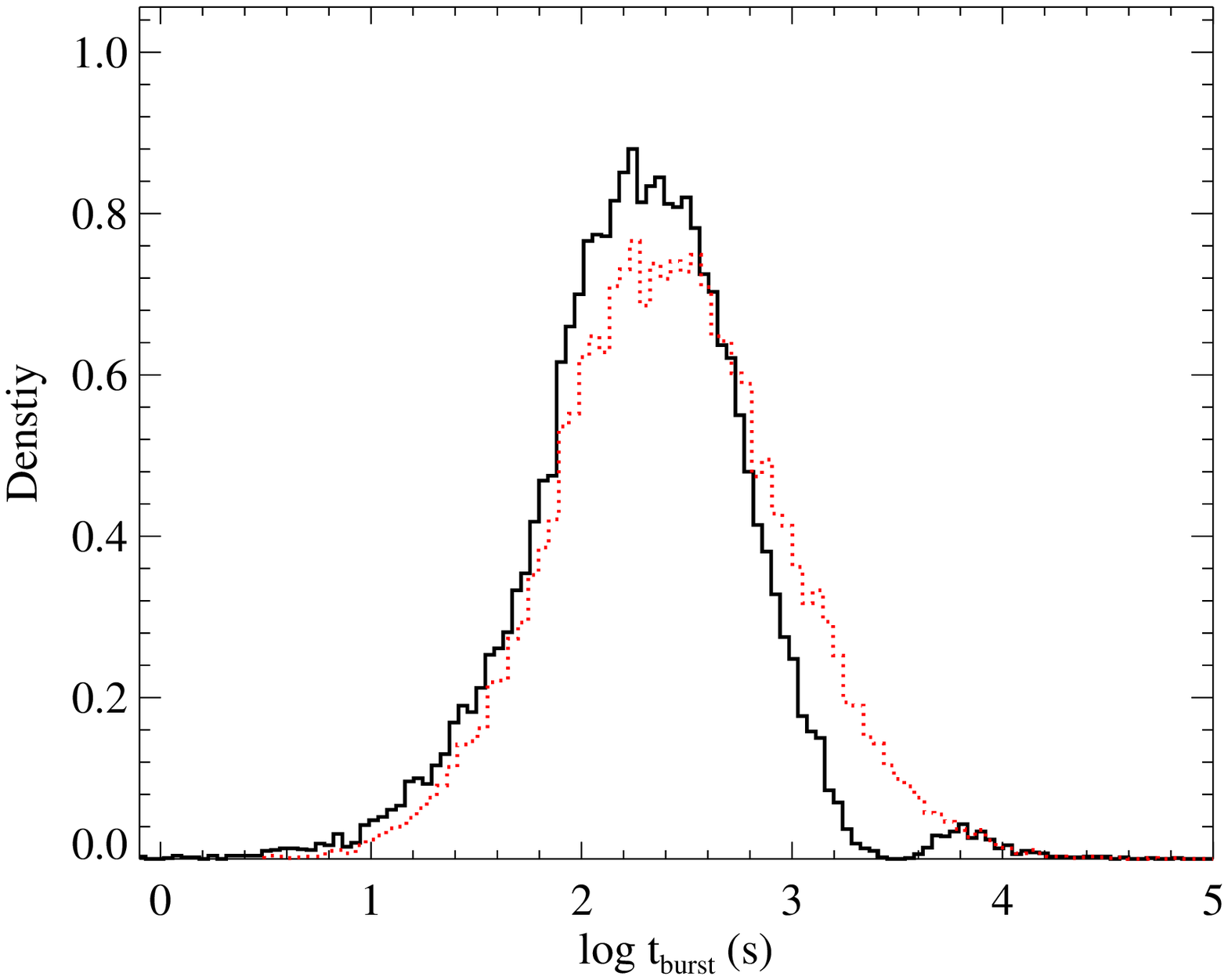} \\
(c) & (d)\\

\\
\end{tabular}

\end{center}
 \caption{
(a) Assumed intrinsic $t_{\rm burst,int}$ distribution, which is a Gaussian distribution in log scale with a mean value $\mu=2.2$ and a standard deviation $\sigma$=0.6; (b) distribution of the observed $T_{\rm 90}$ of the 647 GRBs in the full sample; (c) distributions of the observed $t_{gap,1}$, $t_{gap,2}$; (d) distribution of the simulated ``observed'' value $t_{burst}$. The intrinsic distribution is also plotted as the red dotted histogram for comparison.
 }
\label{fig:eps_e}
\end{figure}

\begin{figure}[ht!]

\begin{center}
\begin{tabular} {c}
\includegraphics[scale=0.38]{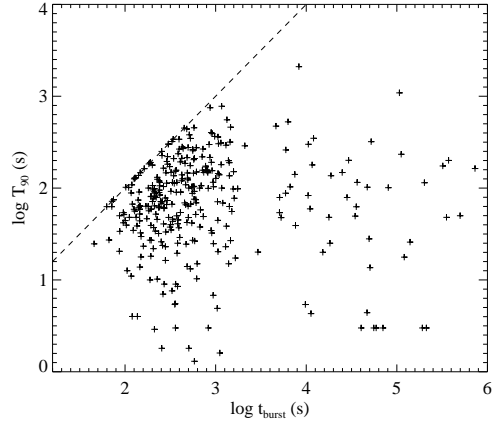}\\
\end{tabular}

\end{center}
 \caption{% 
$T_{90}$ vs $t_{burst}$ for all the bursts in our sample. The dashed line marks where $T_{90}$=$t_{\rm burst}$.
}
\label{fig:eps_e}
\end{figure}

\begin{figure}[ht!]

\begin{center}
\begin{tabular} {c}
\includegraphics[scale=0.38]{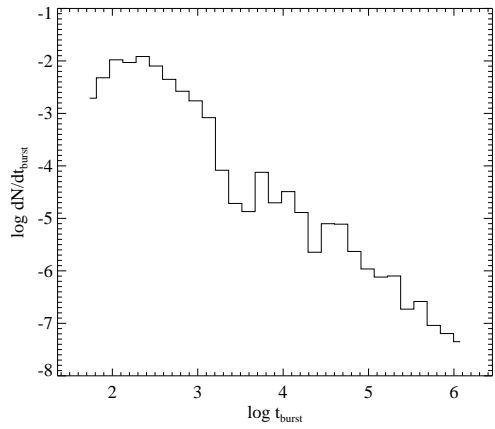}\\
\end{tabular}

\end{center}
 \caption{% 
The $dN/dt_{\rm burst}$ diagram, which does not show an apparent plateau as
suggested by Bromberg et al. (2013).}
\label{fig:eps_e}
\end{figure}

\clearpage
\LongTables
\begin{deluxetable}{ll|ll|ll|ll|ll} 
\tabletypesize{\tiny}
\tablecolumns{10} 
\tablecaption{$t_{\rm burst}$ of the each GRB in our good sample} 
\tablehead{
\colhead{GRB} & \colhead{log $t_{\rm burst}$} &\colhead{GRB} & \colhead{log $t_{\rm burst}$} &\colhead{GRB} & \colhead{log $t_{\rm burst}$} &\colhead{GRB} & \colhead{log $t_{\rm burst}$} &\colhead{GRB} & \colhead{log $t_{\rm burst}$} \\
\hline \\
\colhead{} & \colhead{[s]} &\colhead{} & \colhead{[s]} &\colhead{} & \colhead{[s]} &\colhead{} & \colhead{[s]} &\colhead{} & \colhead{[s]} 
\\}
\startdata 
140114A&2.846$\pm$0.015&
140108A&2.119$\pm$0.009&
140102A& $\sim$1.798($T_{90}$) &
131127A&2.574$\pm$0.027&
131117A&2.475$\pm$0.048\\
131105A&2.527$\pm$0.014&
131103A&3.268$\pm$0.021&
131030A&2.377$\pm$0.003&
131024B&2.519$\pm$0.086&
131018A&2.463$\pm$0.022\\
131002B&2.373$\pm$0.019&
131002A&1.939$\pm$0.021&
130925A&4.066$\pm$0.002&
130907A&2.990$\pm$0.004&
130831B&2.935$\pm$0.027\\
130831A&2.221$\pm$0.050&
130807A&3.596$\pm$0.041&
130803A&2.155$\pm$0.029&
130722A&2.624$\pm$0.005&
130716A&2.175$\pm$0.142\\
130615A&3.175$\pm$0.044&
130612A&2.032$\pm$0.039&
130609B&2.625$\pm$0.005&
130609A&2.121$\pm$0.068&
130608A&2.774$\pm$0.033\\
130606A&2.697$\pm$0.008&
130605A&2.023$\pm$0.037&
130529A& $\sim$2.107($T_{90}$) &
130528A&3.147$\pm$0.016&
130527A&2.407$\pm$0.024\\
130514A&2.744$\pm$0.016&
130505A&2.509$\pm$0.008&
130427B&2.288$\pm$0.017&
130427A& $\sim$2.212($T_{90}$) &
130418A& $\sim$2.477($T_{90}$) \\
130408A&4.694$\pm$0.039&
130327A&2.422$\pm$0.044&
130315A&3.618$\pm$0.162&
130211A&2.580$\pm$0.019&
130131B&2.481$\pm$0.045\\
130131A&2.780$\pm$0.025&
121229A&2.823$\pm$0.010&
121217A&3.066$\pm$0.004&
121212A&3.018$\pm$0.029&
121211A&2.465$\pm$0.016\\
121128A&2.204$\pm$0.015&
121125A&2.138$\pm$0.045&
121123A&2.979$\pm$0.028&
121108A&2.375$\pm$0.016&
121102A&2.016$\pm$0.042\\
121031A&2.374$\pm$0.028&
121027A&4.549$\pm$0.020&
121024A&2.510$\pm$0.028&
121001A& $\sim$2.167($T_{90}$) &
120922A&2.868$\pm$0.037\\
120811C&2.306$\pm$0.021&
120804A&2.019$\pm$0.048&
120729A& $\sim$1.854($T_{90}$) &
120728A&3.022$\pm$0.051&
120724A&2.282$\pm$0.085\\
120703A&2.007$\pm$0.042&
120701A&2.731$\pm$0.046&
120612A&3.777$\pm$0.035&
120521C&2.576$\pm$0.051&
120521B&2.440$\pm$0.037\\
120521A& $\ge$2.513&
120514A&2.409$\pm$0.009&
120422A&2.701$\pm$0.050&
120401A&3.183$\pm$0.082&
120328A&2.191$\pm$0.016\\
120327A&2.238$\pm$0.041&
120326A&2.429$\pm$0.020&
120324A&2.377$\pm$0.012&
120320A& $\ge$5.146&
120308A&4.555$\pm$0.151\\
120219A&2.756$\pm$0.053&
120215A&2.604$\pm$0.068&
120213A&2.436$\pm$0.030&
120211A&2.381$\pm$0.099&
120119A&4.478$\pm$0.031\\
120118B&2.414$\pm$0.036&
120116A&2.445$\pm$0.026&
120106A&2.151$\pm$0.037&
111229A& $\ge$4.266&
111228A&2.571$\pm$0.055\\
111225A& $\sim$2.029($T_{90}$) &
111215A&3.165$\pm$0.005&
111209A&4.801$\pm$0.025&
111208A& $\ge$4.606&
111123A&2.937$\pm$0.011\\
111121A& $\sim$2.076($T_{90}$) &
111107A&2.769$\pm$0.045&
111103B&2.562$\pm$0.003&
111022B&2.609$\pm$0.049&
111016A&3.790$\pm$0.029\\
111008A&2.475$\pm$0.023&
110921A&2.957$\pm$0.029&
110915A&2.784$\pm$0.014&
110820A&2.747$\pm$0.043&
110818A&3.243$\pm$0.032\\
110808A&2.699$\pm$0.039&
110801A&2.902$\pm$0.027&
110726A&2.338$\pm$0.040&
110709A&2.001$\pm$0.011&
110709B&3.179$\pm$0.002\\
110420A&2.329$\pm$0.024&
110414A&2.871$\pm$0.034&
110411A&2.307$\pm$0.016&
110407A&3.024$\pm$0.044&
110319A&2.167$\pm$0.024\\
110312A&2.508$\pm$0.041&
110223B&3.860$\pm$0.024&
110213A&2.122$\pm$0.017&
110210A&2.953$\pm$0.075&
110205A&2.861$\pm$0.008\\
110119A&2.677$\pm$0.006&
110102A&2.735$\pm$0.024&
101225A& $\ge$5.028&
101219A&2.455$\pm$0.079&
101213A& $\sim$2.130($T_{90}$) \\
101030A&2.735$\pm$0.031&
101023A&2.240$\pm$0.007&
101017A&2.824$\pm$0.029&
101011A& $\sim$1.854($T_{90}$) &
100915A& $\sim$2.301($T_{90}$) \\
100906A& $\ge$5.304&
100905A&2.900$\pm$0.030&
100902A& $\ge$6.173&
100901A&2.771$\pm$0.029&
100823A&2.232$\pm$0.065\\
100816A&2.351$\pm$0.047&
100814A&2.738$\pm$0.016&
100807A&2.419$\pm$0.042&
100805A&2.543$\pm$0.014&
100802A&3.666$\pm$0.102\\
100728A&2.969$\pm$0.015&
100727A&2.742$\pm$0.011&
100725B&2.779$\pm$0.021&
100725A& $\sim$2.149($T_{90}$) &
100704A&2.665$\pm$0.014\\
100621A&2.503$\pm$0.013&
100619A&3.194$\pm$0.008&
100615A&2.134$\pm$0.075&
100614A&2.795$\pm$0.065&
100606A& $\sim$2.681($T_{90}$) \\
100526A&2.737$\pm$0.023&
100522A&2.055$\pm$0.083&
100514A&2.646$\pm$0.024&
100513A&2.760$\pm$0.037&
100504A&2.702$\pm$0.023\\
100425A&2.672$\pm$0.034&
100420A&2.661$\pm$0.118&
100418A&2.500$\pm$0.033&
100413A&2.490$\pm$0.017&
100316D& $\sim$3.114($T_{90}$) \\
100305A&2.389$\pm$0.014&
100302A&3.110$\pm$0.022&
100219A& $\ge$5.070&
100212A&2.876$\pm$0.008&
100205A& $\ge$3.115\\
100117A& $\ge$3.222&
091221&2.398$\pm$0.043&
091130B&2.335$\pm$0.028&
091127&3.745$\pm$0.700&
091104&2.918$\pm$0.059\\
091029&2.279$\pm$0.036&
091026&2.737$\pm$0.010&
091020&2.069$\pm$0.025&
090929B& $\sim$2.556($T_{90}$) &
090926B& $\sim$2.040($T_{90}$) \\
090926A&4.714$\pm$0.018&
090912&2.988$\pm$0.039&
090904B&2.162$\pm$0.027&
090904A&3.002$\pm$0.024&
090812&2.518$\pm$0.012\\
090809&3.993$\pm$0.036&
090807&3.875$\pm$0.020&
090728&2.272$\pm$0.072&
090727& $\sim$2.480($T_{90}$) &
090715B&2.671$\pm$0.005\\
090709A&2.105$\pm$0.013&
090621A&2.851$\pm$0.046&
090618&2.481$\pm$0.008&
090530&2.117$\pm$0.043&
090529&3.067$\pm$0.038\\
090519&2.729$\pm$0.047&
090516&2.764$\pm$0.014&
090515& $\ge$2.454&
090429A& $\sim$2.274($T_{90}$) &
090424&2.016$\pm$0.019\\
090423&2.791$\pm$0.022&
090419& $\sim$2.653($T_{90}$) &
090418A&2.069$\pm$0.017&
090417B&3.322$\pm$0.016&
090407&2.996$\pm$0.025\\
090404&2.384$\pm$0.013&
090401B& $\sim$2.263($T_{90}$) &
090313&4.448$\pm$0.010&
090123& $\sim$2.117($T_{90}$) &
090111&2.975$\pm$0.042\\
081230&2.419$\pm$0.020&
081222&3.038$\pm$0.020&
081221&2.271$\pm$0.009&
081210&2.703$\pm$0.024&
081203A& $\sim$2.468($T_{90}$) \\
081128&2.688$\pm$0.029&
081127&2.567$\pm$0.020&
081118&2.971$\pm$0.045&
081109& $\sim$2.279($T_{90}$) &
081102&3.151$\pm$0.013\\
081028&3.807$\pm$0.016&
081024& $\ge$2.383&
081008&2.642$\pm$0.009&
081007&2.315$\pm$0.039&
080928&2.635$\pm$0.004\\
080919& $\ge$2.852&
080916A&2.232$\pm$0.069&
080906&2.913$\pm$0.023&
080905B&2.244$\pm$0.024&
080810&2.507$\pm$0.013\\
080805&2.444$\pm$0.036&
080727A& $\ge$3.017&
080721&5.214$\pm$0.049&
080707&2.238$\pm$0.039&
080613B&2.412$\pm$0.013\\
080607&2.309$\pm$0.004&
080603B&2.164$\pm$0.021&
080602&2.146$\pm$0.021&
080523& $\sim$2.009($T_{90}$) &
080506&2.790$\pm$0.014\\
080503& $\ge$2.888&
080413A&2.208$\pm$0.039&
080328&2.191$\pm$0.016&
080325&2.689$\pm$0.148&
080320&2.685$\pm$0.020\\
080319D&2.957$\pm$0.028&
080319A& $\ge$4.894&
080310&4.966$\pm$0.043&
080307& $\sim$2.100($T_{90}$) &
080229A&2.293$\pm$0.008\\
080212&2.693$\pm$0.005&
080210& $\ge$5.031&
080207& $\sim$2.531($T_{90}$) &
080205&2.267$\pm$0.016&
080123&2.572$\pm$0.031\\
080120& $\ge$4.183&
071227&2.704$\pm$0.053&
071118&2.958$\pm$0.024&
071112C&3.082$\pm$0.041&
071031&3.062$\pm$0.025\\
071028A&2.752$\pm$0.044&
070808&2.327$\pm$0.061&
070724A&2.528$\pm$0.060&
070721B&2.594$\pm$0.005&
070704&2.719$\pm$0.036\\
070621&2.583$\pm$0.033&
070616&3.078$\pm$0.083&
070611&3.633$\pm$0.035&
070529&2.219$\pm$0.018&
070520B&2.666$\pm$0.024\\
070520A&2.297$\pm$0.077&
070518&2.553$\pm$0.039&
070429A&2.818$\pm$0.021&
070420&2.306$\pm$0.013&
070419B&2.588$\pm$0.017\\
070419A&2.846$\pm$0.067&
070412&1.942$\pm$0.041&
070318& $\sim$1.873($T_{90}$) &
070311& $\ge$5.689&
070306&2.565$\pm$0.027\\
070224&2.950$\pm$0.065&
070220& $\sim$2.111($T_{90}$) &
070208&3.722$\pm$0.042&
070129&3.168$\pm$0.020&
070110&4.535$\pm$0.036\\
070107&2.721$\pm$0.010&
061222B&2.619$\pm$0.087&
061222A&2.331$\pm$0.014&
061202&2.605$\pm$0.027&
061121&2.328$\pm$0.014\\
061110A&2.747$\pm$0.107&
061102&2.269$\pm$0.084&
061028&2.817$\pm$0.037&
061006&2.548$\pm$0.112&
060929&3.141$\pm$0.032\\
060906&2.525$\pm$0.065&
060904B&2.495$\pm$0.006&
060814&2.856$\pm$0.055&
060801& $\ge$2.754&
060729&4.569$\pm$0.006\\
060719&2.080$\pm$0.025&
060714&2.453$\pm$0.023&
060708&2.356$\pm$0.034&
060614&2.667$\pm$0.032&
060607A&4.673$\pm$0.159\\
060604&2.380$\pm$0.005&
060526& $\ge$5.497&
060522&2.404$\pm$0.033&
060512&2.580$\pm$0.042&
060510B&2.767$\pm$0.010\\
060510A&2.124$\pm$0.034&
060502A&2.324$\pm$0.034&
060428B&2.833$\pm$0.022&
060428A&2.014$\pm$0.038&
060418&2.294$\pm$0.012\\
060413&3.022$\pm$0.030&
060306&2.193$\pm$0.054&
060219&2.348$\pm$0.028&
060218&4.073$\pm$0.017&
060211A&2.665$\pm$0.080\\
060210&2.644$\pm$0.007&
060204B&2.626$\pm$0.007&
060202&3.096$\pm$0.028&
060124&2.980$\pm$0.001&
060115&3.011$\pm$0.041\\
060111B&2.141$\pm$0.031&
060111A&2.721$\pm$0.018&
060109&2.382$\pm$0.036&
051210& $\ge$2.750&
051117A&4.358$\pm$0.023\\
051016B&2.132$\pm$0.039&
051016A&2.446$\pm$0.052&
051001&3.129$\pm$0.031&
050922C&2.689$\pm$0.029&
050922B&3.229$\pm$0.012\\
050915B&2.637$\pm$0.023&
050915A&2.339$\pm$0.046&
050904& $\ge$5.498&
050822&2.944$\pm$0.025&
050819&2.827$\pm$0.046\\
050814&2.957$\pm$0.022&
050803&3.772$\pm$0.010&
050730&2.853$\pm$0.011&
050726&4.061$\pm$0.040&
050724&2.895$\pm$0.020\\
050716&2.819$\pm$0.024&
050713B&2.604$\pm$0.160&
050713A&2.506$\pm$0.027&
050502B& $\ge$5.427&
050421& $\ge$2.796\\
050406&2.560$\pm$0.044&
050319&2.555$\pm$0.029&
050315&2.306$\pm$0.081&
\enddata 

\end{deluxetable}

\end{document}